\documentclass[PAPER, cernpreprint=true, atlasdraft=false, coverpage=false, UKenglish, texmf, orcidlogo]{atlasdoc}
\usepackage[subcaption=true, subfigure=false, subfig=false]{atlaspackage}
\usepackage{multirow}
\usepackage{makecell}
\usepackage{atlasbiblatex}

\usepackage{atlasphysics}
\graphicspath{{logos/}{figures/}}

\usepackage{ANA-EXOT-2022-31-PAPER-defs}
\AtlasTitle{Search for emerging jets in $pp$ collisions  at $\sqrt{s} = 13.6$~\TeV with the ATLAS experiment}

\AtlasAbstract{%
A search for emerging jets is presented using 51.8~fb$^{-1}$ of
proton--proton collision data at $\sqrt{s} = 13.6$~\TeV, collected by the
ATLAS experiment during 2022 and 2023. The search explores a hypothetical
dark sector featuring “dark quarks” that are charged under a confining
gauge group and couple to the Standard Model via a new mediator particle.
These dark quarks undergo showering and hadronisation within the dark
sector, forming long-lived dark mesons that decay back into Standard Model
particles. This results in jets that contain multiple displaced vertices known as emerging jets. The analysis targets events with pairs of
emerging jets, produced either through a vector mediator, $Z'$, in the
$s$-channel, or a scalar mediator, $\Phi$, in the $t$-channel. No significant
excess over the Standard Model background is observed. Assuming a dark pion
proper decay length between 5~mm and 50~mm, $Z'$ mediator masses between 600~\GeV
and 2550~\GeV are excluded for quark and dark quark coupling values of 0.01
and 0.1, respectively. For a quark dark-quark coupling of $0.1$, $\Phi$
mediator masses between 600~\GeV and 1375~\GeV are excluded. These results
represent the first direct search targeting emerging jet pair production via
a $Z'$ mediator, as well as the first study of emerging jet production
mediated by a scalar particle exchanged in the $t$-channel.
}

\AtlasRefCode{EXOT-2022-31}

\PreprintIdNumber{CERN-EP-2025-099}

\AtlasJournal{Rep. Prog. Phys.}
\AtlasJournalRef{Rep. Prog. Phys. 88 (2025) 097801}
\AtlasDOI{10.1088/1361-6633/adfe17}

\AtlasCoverEgroupAnalysisTeam{atlas-exot-2022-31-analysis-team@cern.ch}

\hypersetup{pdftitle={ATLAS document},pdfauthor={The ATLAS Collaboration}}

\begin{document}

\maketitle

\tableofcontents

%
%

%
\section{Introduction}
\label{sec:intro}

Dark Matter (DM) remains one of the most significant mysteries in physics.
Despite overwhelming evidence for the existence of DM~\cite{Rubin:1980zd,Persic:1995ru,Clowe:2006eq}, its
particle nature and interactions with Standard Model (SM)
fields have yet to be determined.
Various extensions to the SM have proposed new stable,
weakly interacting massive particles (WIMPs) as DM candidates, motivated
by their natural production mechanism in the early universe through
thermal freeze-out~\cite{Jungman:1995df,Steigman:2012nb}.
However, decades of experimental searches, including direct detection~\cite{PhysRevLett.119.181302,PhysRevD.100.022001,PhysRevD.98.102006,PhysRevD.100.102002,PhysRevLett.121.081307,PhysRevLett.118.021303,PhysRevLett.126.091301,Lai_2023,PhysRevD.99.062001},
indirect detection~\cite{Ackermann_2015,HESS:2022ygk}, and collider experiments~\cite{EXOT-2018-06,CMS-EXO-20-004}, have increasingly
constrained the parameter space for WIMPs, prompting the exploration of
alternative theoretical frameworks that can address the DM puzzle.

One particularly compelling alternative to the WIMP paradigm is the
existence of a ``dark sector'' (DS), consisting of particles that are
neutral under the SM gauge group but charged under a new symmetry group within
the DS~\cite{Strassler:2006im,Strassler:2006ri,Han:2007ae}. In these
models, interactions between DS states and SM particles occur via
mediator particles that couple to both sectors. Among the various
possibilities, dark sectors with a gauge structure and particle content
resembling quantum chromodynamics (QCD) have attracted significant
theoretical interest due to their rich phenomenology and their potential
to naturally explain the observed relic abundance of
DM~\cite{Bai:2013xga,Beauchesne:2018myj,Albouy:2022cin}.
In these ``dark QCD'' models,
the SM is extended with $N_f$ flavours of dark quarks that are charged under a new non-Abelian gauge symmetry, $\mathrm{SU}(N_c)$, which
confines at a scale $\Lambda_D$, where $N_c$ is the number of dark colours.
As in QCD, these dark quarks undergo
parton showering and subsequently hadronise, forming bound dark hadron
states. While some dark hadrons may be stable, providing viable DM
candidates, others can be unstable and decay back into SM particles via an off-shell exchange of the mediator particle.

The lifetimes of unstable dark mesons depend on the strength of their effective couplings to
the SM\@. For large couplings, dark mesons will decay promptly, producing ``dark jets'' with distinctive
substructure features or, in cases where some dark states remain invisible, giving rise to “semi-visible jets” with significant missing transverse momentum aligned with the jet direction~\cite{Cohen_2015}. However,
for weaker couplings, unstable dark mesons naturally acquire longer lifetimes, leading to the
distinctive topology of an ``emerging jet'' (EJ)~\cite{Schwaller:2015gea}. These jets are characterised by multiple secondary vertices within the jet cone, with potentially macroscopic displacements from the primary $pp$ interaction point, resulting from the gradual emergence of SM particles as the dark mesons decay in flight.

The ATLAS and CMS collaborations have carried out multiple searches for
signatures predicted by dark QCD scenarios, focusing on a variety of production
mechanisms and jet topologies. ATLAS has investigated the resonant production of
dark jets, where all dark mesons decay promptly into SM
particles~\cite{HDBS-2018-45}.
Semi-visible jet scenarios have been studied by both ATLAS and
CMS in the $s$-channel~\cite{CMS-EXO-19-020,EXOT-2021-19} and $t$-channel~\cite{EXOT-2022-37}. A Run~1 ATLAS search explored
scenarios with long-lived dark pions decaying within the tracker volume, giving
rise to signatures resembling emerging jets~\cite{EXOT-2013-12}. Additionally, CMS has
conducted two dedicated searches for emerging jets, both considering the pair
production of a bi-fundamental mediator that is charged under both the dark and
SM gauge groups, which placed limits on mediator masses up to 1.9 \TeV, assuming
a dark pion proper decay length of approximately
100~mm~\cite{CMS-EXO-18-001,CMS-EXO-22-015}. However, these searches have
exclusively focused on the pair production of a bi-fundamental mediator, leaving
alternative production mechanisms, such as $s$-channel exchange of a vector
mediator, largely unexplored.

\begin{figure}[!htbp]
\centering
\includegraphics[width=0.4\textwidth]{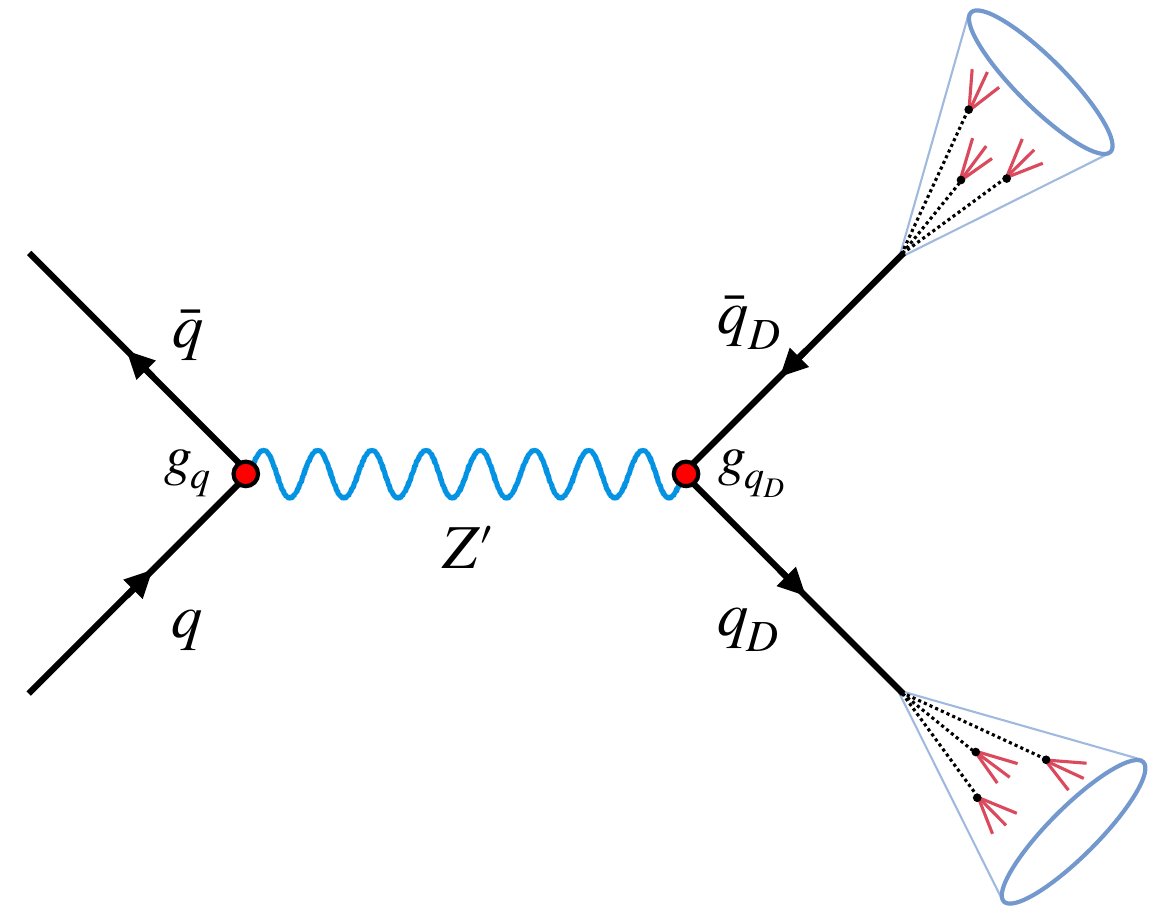}
\caption{A diagram illustrating the production of emerging jets via an
$s$-channel mediator, $Z'$. Filled circles represent the $Z'$ couplings. Emerging jets are depicted as cones, with the trajectories of long-lived dark pions shown as dashed lines and solid lines indicating their Standard Model decay products.}
\label{fig:feynman_EJ}
\end{figure}

The search presented in this paper considers emerging jet production from two
different mediator scenarios. The first scenario involves $s$-channel production
through a new vector boson $Z'$ that couples both to SM
quarks and to a dark sector containing dark quarks.
The vector portal is one of the most widely studied mechanisms for connecting
the Standard Model to a dark sector, appearing in a broad class of dark matter
and hidden valley
models~\cite{Englert:2016joy,Bernreuther:2019pfb,Cheng:2019yai}.
As illustrated in
Figure~\ref{fig:feynman_EJ}, the $Z'$ is produced in proton--proton ($pp$) collisions via
quark--antiquark annihilation, $q\bar{q} \to Z'$, and subsequently decays into a pair of
dark quarks, $Z' \to q_D \bar{q}_D$. These dark quarks then undergo showering and
hadronisation within the dark sector, leading to the formation of
emerging jets.

The interactions of the $Z'$ mediator with both the SM and DS are
governed by two distinct couplings. The coupling between the $Z'$ and the SM quarks
is denoted by $g_q$, which determines the production cross-section of the mediator. The coupling between the
$Z'$ and the dark quarks is denoted by $g_{q_D}$, which, along with $g_q$ controls the branching fraction of the
mediator’s decay into the dark sector.

\begin{figure}[h]
\centering
\begin{subfigure}[b]{0.31\textwidth}
\centering
\includegraphics[width=\textwidth]{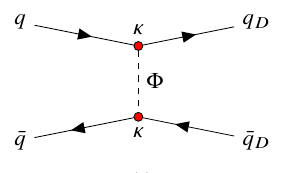}
\caption{}
\end{subfigure}
\qquad
\begin{subfigure}[b]{0.43\textwidth}
\centering
\includegraphics[width=\textwidth]{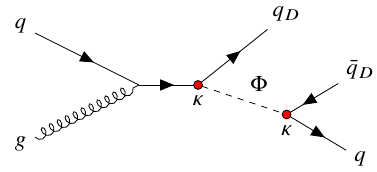}
\caption{}
\end{subfigure} \\
\begin{subfigure}[b]{0.34\textwidth}
\centering
\includegraphics[width=\textwidth]{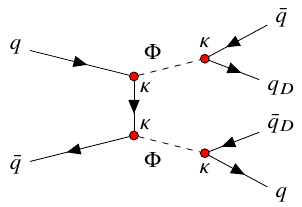}
\caption{}
\end{subfigure}
\caption{Feynman diagrams for $q_D \bar{q}_D$ production with up to two
additional SM jets via a scalar $\Phi$.}
\label{fig:feynman_tchannel}
\end{figure}

In the second scenario, emerging jets are produced via a $t$-channel
process, where $\Phi$, a bi-fundamental scalar mediator, serves as a
portal between the SM and the DS.
In this model, $\Phi$ couples to both down-type SM quarks and dark quarks
via a coupling matrix $\kappa$, as described in Ref.~\cite{Carmona:2024tkg}.
As illustrated in
Figure~\ref{fig:feynman_tchannel}, the final state consists of two
dark quarks and up to two SM quarks, resulting in a signature similar
to the $s$-channel case, characterised by the presence of two emerging
jets.

This analysis uses two complementary strategies to identify emerging jet
signatures: one using selections on high-level jet observables and another using
machine learning (ML) techniques to improve signal identification. To maximize sensitivity across different mediator
mass ranges, the search is divided into two regions based on the invariant mass of the two leading jets ($m_{jj}$), each with distinct trigger strategies. The \lowmjj region ($m_{jj} < 1$~\TeV) utilises a dedicated jet trigger, introduced for Run 3, specifically designed to select emerging jets. The \highmjj region ($m_{jj} > 1$~\TeV) relies on a standard single-jet trigger. Combining these two regions provides sensitivity for
mediator masses between 600~\GeV and 3500~\GeV.
The dominant background comes from QCD multijet
events, which are modelled using a data-driven approach.

This paper is structured as follows: Section~\ref{sec:atlas} provides a
description of the ATLAS detector, followed by an overview of the dataset and
simulated samples used in the analysis in Section~\ref{sec:simulation}. The
definitions of the reconstructed objects are outlined in
Section~\ref{sec:objects}. Section~\ref{sec:analysis} details the analysis
strategies and event selection criteria for each signal region.
Background estimation strategies are described in
Section~\ref{sec:background}, while systematic uncertainties are discussed in
Section~\ref{sec:systematics}. Finally, the statistical analysis and results
are presented in Section~\ref{sec:results}, with conclusions provided in
Section~\ref{sec:conclusion}.



%
\section{ATLAS detector}
\label{sec:atlas}

The ATLAS detector~\cite{PERF-2007-01,GENR-2019-02} at the LHC covers nearly the entire solid angle around the collision point.\footnote{\AtlasCoordFootnote}
It consists of an inner tracking detector surrounded by a thin superconducting solenoid, electromagnetic and hadronic calorimeters,
and a muon spectrometer incorporating three large superconducting air-core toroidal magnets.

The inner-detector system (ID) is immersed in a \SI{2}{\tesla} axial magnetic field
and provides charged-particle tracking in the range \(|\eta| < 2.5\).
The high-granularity silicon pixel detector covers the vertex region and typically provides four measurements per track,
the first hit normally being in the insertable B-layer installed before Run~2~\cite{ATLAS-TDR-19,PIX-2018-001}.  The geometry of the pixel detector consists of concentric barrel layers with $r= 33$, 50.5, 88.5, and 122.5~mm in the central region, and three disks in each of the endcaps at $|z| = 495$, 580, and 650~mm.
It is followed by the silicon microstrip tracker (SCT), consisting of barrel layers at $r=299$, 371, 443, and 514~mm, spanning $|z| < 746$~mm, and nine wheels in each of the endcaps with $854 < |z| < 2720$~mm.  The SCT usually provides eight measurements per track.
These silicon detectors are complemented by the transition radiation tracker (TRT),
which enables radially extended track reconstruction up to \(|\eta| = 2.0\).
The TRT also provides electron identification information
based on the fraction of hits (typically 30 in total) above a higher energy-deposit threshold corresponding to transition radiation.

The calorimeter system covers the pseudorapidity range \(|\eta| < 4.9\).
Within the region \(|\eta|< 3.2\), electromagnetic calorimetry is provided by barrel and
endcap high-granularity lead/liquid-argon (LAr) calorimeters,
with an additional thin LAr presampler covering \(|\eta| < 1.8\)
to correct for energy loss in material upstream of the calorimeters.
Hadronic calorimetry is provided by the steel/scintillator-tile calorimeter,
segmented into three barrel structures within \(|\eta| < 1.7\), and two copper/LAr hadronic endcap calorimeters.
The solid angle coverage is completed with forward copper/LAr and tungsten/LAr calorimeter modules
optimised for electromagnetic and hadronic energy measurements, respectively.

The muon spectrometer comprises separate trigger and
high-precision tracking chambers measuring the deflection of muons in a magnetic field generated by the superconducting air-core toroidal magnets.
The field integral of the toroids ranges between \num{2.0} and \qty{6.0}{\tesla\metre}
across most of the detector.
Three layers of precision chambers, each consisting of layers of monitored drift tubes, cover the region \(|\eta| < 2.7\),
except in the innermost layer of the endcap region, where layers of small-strip thin-gap chambers and Micromegas chambers both provide precision tracking in the region \(1.3 < |\eta| < 2.7\).
The muon trigger system covers the range \(|\eta| < 2.4\) with resistive-plate chambers in the barrel, thin-gap chambers in the endcap regions, and the aforementioned small-strip thin-gap chambers and Micromegas chambers in the innermost layer of the endcap.

The luminosity is measured mainly by the LUCID--2 detector that records Cherenkov light produced in the quartz windows of photomultipliers located close to the beampipe~\cite{LUCID2}.

Events are selected by the first-level trigger system implemented in custom hardware,
followed by selections made by algorithms implemented in software in the high-level trigger~\cite{TRIG-2022-01}.
The first-level trigger accepts events from the \qty{40}{\MHz} bunch crossings at a rate below \qty{100}{\kHz},
which the high-level trigger further reduces in order to record complete events to disk at an average rate of about \qty{3}{\kHz}.

The Run~3 detector configuration benefits from several upgrades compared with that of Run~2 to maintain high detector performance at the higher pile-up levels of Run~3. The improvements include a new innermost layer of the muon spectrometer in the endcap region, which provides higher redundancy and a large reduction in fake muon triggers. The trigger system also benefits from new LAr digital electronics with significantly increased granularity. Other updates and further details are provided in Ref.~\cite{GENR-2019-02}.

A software suite~\cite{mySOFT-2022-02} is used in data simulation, in the reconstruction
and analysis of real and simulated data, in detector operations, and in the trigger and data acquisition
systems of the experiment.


%
\section{Data and simulated events}
\label{sec:simulation}

This analysis uses 51.8~fb$^{-1}$ of $\sqrt{s} = 13.6$~\TeV $pp$ collision data
collected by the ATLAS experiment from 2022 to 2023 during Run~3 of the LHC\@.
Only data recorded under stable beam conditions with all detector subsystems fully operational are included, and standard data quality requirements are applied to ensure reliable event reconstruction~\cite{DAPR-2018-01}.

Monte Carlo (MC) events are used to evaluate the signal acceptance, optimise the
analysis methodology, and validate the performance of the background estimation
methods. The effect of multiple interactions in the same and neighbouring bunch
crossings (pile-up) is modelled by overlaying~\cite{SIMU-2020-01} the simulated
hard-scattering event with inelastic \(pp\) events generated
from a mix of \EPOS[2.0.1.4]~\cite{myWerner:2005jf} and
\PYTHIA[8.308]~\cite{Bierlich:2022pfr}. The MC events are then weighted to
reproduce the distribution of the average number of interactions per bunch
crossing ($\left<\mu \right>$) observed in the data. All simulated events are
processed through a detailed simulation of the ATLAS detector using
$\GEANT$~\cite{Agostinelli:2002hh} to model the detector
response~\cite{SOFT-2010-01}, and are reconstructed using the same algorithms
applied to the data.

\subsection{Signal simulation}

Samples of $pp\rightarrow Z'\rightarrow q_D\bar{q}_D$ are simulated using
\PYTHIA[8.309]~\cite{Bierlich:2022pfr} with $Z'$ masses ranging from 600~\GeV to
3.5~\TeV. The production cross-section is a free parameter that depends on the
coupling of the $Z'$ to SM quarks, $g_q$. For benchmark values of $g_q = 0.01$
and $g_{q_D} = 0.1$, the production cross-section computed at leading-order (LO) ranges from
$220~\mathrm{fb}$ at $m_{Z'} = 600$~\GeV to $0.035~\mathrm{fb}$ at
$m_{Z'} = 3500$~\GeV, with branching ratios to dark quarks of approximately 85\%.
For the cross-section values explored, varying $g_q$ does not affect the dijet
mass distribution because the true $Z'$ width remains much smaller than the
reconstructed width, which is dominated by resolution and reconstruction
effects.
A fixed width of 10 GeV is used for all $Z'$ samples.
When calculating the theoretical cross-sections it is assumed that the
$Z'$ mediator has axial-vector couplings~\cite{Abercrombie:2015wmb}.

Signal samples for $t$-channel production are generated using the
\MGNLO v3.4.2 event generator~\cite{Alwall:2014hca}, with the
\texttt{\detokenize{darkqcd_fv_down}} model from Ref.~\cite{Carmona:2024tkg} used
to calculate matrix elements (ME) at LO, including up to two
additional partons. The coupling between the mediator $\Phi$, dark quarks, and
SM quarks is described by a matrix $\kappa_{ij}$. In this study, $\Phi$ is
assumed to couple only to down-type SM quarks~\footnote{The choice of coupling to down-type quarks corresponds to the scenario where $\Phi$ is an $SU(2)_L$ singlet with hypercharge $Y = 1/3$. Alternative quantum number assignments could instead lead to couplings to up-type quarks. For the values of $\kappa$ considered in this analysis, this choice has a negligible impact on both the final-state signature and the theoretical production cross-section~\cite{Carmona:2024tkg}.} and to the first-generation dark
quark. As a result, only the matrix elements $\kappa_{ij}$ with $i = 1$
(corresponding to the dark quark flavour) and $j = 1, 2, 3$ (corresponding to
the SM down-type quark flavours) are considered. For $\kappa_{1j} = 0.1$, the
cross-section computed at LO ranges from
$10~\mathrm{fb}$ at $m_{\Phi} = 600$~\GeV to $0.0082~\mathrm{fb}$ at
$m_{\Phi} = 2000$~\GeV.
Jet matching is performed using the MLM~\cite{Mangano:2006rw}
scheme, with a matching parameter of 20~\GeV. For both mediator scenarios, the
\NNPDF[2.3lo]~\cite{Ball:2012cx} parton distribution function (PDF) set and the A14
tuned parameter set~\cite{ATL-PHYS-PUB-2014-021} are used.

In all samples, the dark sector showers are produced in \PYTHIA[8.309] using the
Hidden Valley module~\cite{Carloni:2010tw,Carloni:2011kk}.  Following
Ref.~\cite{Schwaller:2015gea}, the number of dark colours is set to $N_c = 3$,
the number of dark quark flavours is set to $N_f = 7$, and the confinement
scale, $\Lambda_D$, dark quark mass, $m_{q_D}$, dark pion mass, $m_{\pi_D}$, and
dark vector meson mass, $m_{\rho_D}$ are chosen to satisfy the following
hierarchy: $m_{\rho_D} = 2 \Lambda_D = 2 m_{q_D} = 4 m_{\pi_D}$.
This mass hierarchy is motivated by QCD, where pions arise as
pseudo-Nambu--Goldstone bosons of spontaneously broken chiral symmetry, implying
that $m_{\pi_D} < \Lambda_D$. The choice of $N_f = 7$ and $N_c = 3$ is motivated
by the results of Ref.~\cite{Bai:2013xga}, which show that these values can
produce a dark matter--to--baryon abundance ratio close to the observed value.
The dark rho mesons are set to decay via $\rho_D \to \pi_D \pi_D$, with
$m_{\rho_D}$ chosen to ensure that this decay is kinematically allowed and
dominant. The resulting dark pions are forced to decay into down quarks.
The proper decay length of the dark pions is fixed per generated sample with
values ranging from 1~mm to 1000~mm, ensuring that the majority of decays occur
prior to reaching the calorimeter.  Dark baryon production, which is suppressed
by a factor of $1/N_c^2$~\cite{Cohen_2015}, is not considered.  Three different
sets of model parameters are considered, corresponding to \( m_{\pi_D} \) values
of 5, 10, and 20~\GeV, respectively. A summary of the signal benchmarks is
provided in Table~\ref{tab:signal_benchmarks}.

\begin{table}[h!]
\begin{center}
\caption{Benchmark models used in the analysis. In the headings, $\Lambda_D$ corresponds to the dark confinement scale and the masses $m_{\pi_D}$ and $m_{\rho_D}$ correspond to the masses of the dark pion and dark rho mesons, respectively. For each mediator mass—either $m_{Z'}$ or $m_{\Phi}$—and each set of dark sector parameters ($m_{\pi_D}$, $\Lambda_D$, $m_{\rho_D}$), samples are generated with dark pion proper decay lengths of 5~mm, 50~mm, and 500~mm, indicated in bold in the table. For mediator masses $m_{Z'}$ and $m_{\Phi}$ equal to 0.6, 1.5, or 3.0~\TeV, an extended grid of dark pion lifetimes is generated, spanning proper decay lengths from 1~mm to 1000~mm.}
\label{tab:signal_benchmarks}
\begin{tabular}{c|c|c|c}
\toprule
($m_{\pi_D}$, $\Lambda_D$, $m_{\rho_D}$)~[$\GeV$]     &  \multicolumn{3}{c}{(5, 10, 20), (10, 20, 40), (20, 40, 80) }\\
\midrule
$m_{Z'}$~[$\TeV$]       & \multicolumn{3}{c}{0.6, 0.8, 1.0, 1.2, 1.5, 1.8, 2.2, 2.6, 3.0, 3.5}\\
$m_{\Phi}$~[$\TeV$]     & \multicolumn{3}{c}{0.6, 1.0, 1.5, 2.0}\\
\midrule
$c\tau_{\pi_D}$~[\text{mm}]  & \multicolumn{3}{c}{1, \textbf{5}, 10, \textbf{50}, 100, \textbf{500}, 1000 }\\
\bottomrule
\end{tabular}
\label{tab:signal}
\end{center}
\end{table}

\subsection{Background simulation}

While the background modelling strategy used in this analysis is fully data driven,
samples of simulated background events are used to optimise the analysis selections
and validate the background estimation methods.

Multijet production was generated using \PYTHIA[8.230]~\cite{Sjostrand:2014zea} with ME at
LO for dijet production which were matched to the parton shower.
The \NNPDF[2.3lo] PDF set was used in
the matrix element generation, the parton shower, and the simulation of the
multi-parton interactions. The A14~\cite{ATL-PHYS-PUB-2014-021}
set of tuned parameters was used.

The production of \ttbar events was modelled using the
\POWHEGBOX[v2]~\cite{Frixione:2007nw,Nason:2004rx,Frixione:2007vw,Alioli:2010xd}
generator at next-to-leading-order (NLO) with the \NNPDF[3.0nlo]~\cite{Ball:2014uwa} PDF set
and the \hdamp parameter\footnote{The
\hdamp parameter is a resummation damping factor and one of the
parameters that controls the matching of \POWHEG ME to
the parton shower and thus effectively regulates the
high-\pT radiation against which the \ttbar system recoils.} set
to 1.5\,\mtop~\cite{ATL-PHYS-PUB-2016-020}.  The events were interfaced
to \PYTHIA[8.230]~\cite{Sjostrand:2014zea} to model the parton shower,
hadronisation, and underlying event, with parameters set according
to the A14 tune~\cite{ATL-PHYS-PUB-2014-021} and using the \NNPDF[2.3lo]
set of PDFs. The decays of bottom and charm hadrons
were performed by \EVTGEN[1.6.0]~\cite{Lange:2001uf}.

The production of \(V+\)jets was simulated with the
\SHERPA[2.2.14]~\cite{Bothmann:2019yzt}
generator using ME at NLO for up to two partons, and LO for up to five partons calculated with the Comix~\cite{Gleisberg:2008fv}
and \OPENLOOPS~\cite{Buccioni:2019sur,Cascioli:2011va,Denner:2016kdg} libraries. They
were matched with the \SHERPA parton shower~\cite{Schumann:2007mg} using the \MEPSatNLO
prescription~\cite{Hoeche:2011fd,Hoeche:2012yf,myCatani:2001cc,Hoeche:2009rj}
using the set of tuned parameters developed by the \SHERPA authors.
The \NNPDF[3.0nnlo] set of PDFs was used and the samples
were normalised to a next-to-next-to-leading-order (NNLO)
prediction~\cite{Anastasiou:2003ds}.


%
\section{Event reconstruction and selections}
\label{sec:objects}

Tracks are reconstructed from collections of energy deposits (hits) in the ID using a combinatorial
Kalman filter~\cite{Fruhwirth:1987fm} in two passes. The primary pass is optimised for
reconstructing tracks originating near the primary interaction point (IP) and is
designed to efficiently reconstruct prompt charged particles with $\pt > 500~\MeV$. This algorithm primarily
selects tracks with transverse impact parameters of $|d_0| < 5$ mm relative to the IP,
making it largely insensitive to displaced tracks from decays of long-lived particles (LLPs).
To extend tracking acceptance to displaced signatures, a secondary large-impact
parameter track reconstruction pass is performed using hits not associated with tracks in the
primary pass. This secondary pass relaxes the impact parameter requirements,
enabling the reconstruction of tracks with $\pt > 1~\GeV$ and transverse impact parameters up to
$|d_0| < 300$~mm~\cite{IDTR-2021-03}.

The jet reconstruction strategy employed in this analysis is tailored to the distinct characteristics of jets originating
from dark sector quarks. Unlike conventional QCD jets, these jets undergo a two-stage showering process—first within
the dark sector and then in the SM—resulting in a broader radiation pattern.
To fully capture the resulting hadronic shower, jets are reconstructed using the anti-$k_t$ algorithm~\cite{Cacciari:2008gp,Fastjet} with a radius parameter of $R = 1.0$.
Additionally, the displaced nature of the jet constituents poses challenges for track-based jet reconstruction.
In particular, the ATLAS particle flow algorithm~\cite{PERF-2015-09}, which associates jet
constituents with the primary hard-scatter vertex (PV), may reject displaced
tracks that fail this association, thereby reducing sensitivity to the emerging jet topology.
To mitigate these effects, a calorimeter-based jet reconstruction strategy is
adopted, where uncalibrated topological clusters~\cite{PERF-2014-07,JETM-2018-05} are first clustered into $R=0.4$ jets using the anti-$k_t$ algorithm. The subset of these small-$R$ jets with transverse momentum $\pt>15$~\GeV are subsequently reclustered into large-$R$ ($R=1.0$) jets. This reclustering approach allows for the calibration and
uncertainty assessment of the small-$R$ jets using standard ATLAS techniques,
which can then be propagated to the large-$R$ jets. Additionally, the
transverse momentum requirement on the small-$R$ jets acts as an effective
grooming mechanism, reducing soft contamination without requiring additional
grooming procedures.
To facilitate the data-driven background estimation strategy outlined in
Section~\ref{sec:background}, small-$R$ jets are identified as $b$-jets if they
satisfy the 77\% efficiency working point of the DL1r algorithm~\cite{FTAG-2019-07}.
As part of the reconstruction of reclustered $R=1.0$ jets, a track-to-jet
association is performed for both standard and large-impact parameter tracks using the ghost-association
technique~\cite{Cacciari:2008gn} that defines which tracks are later used to calculate discriminants for emerging jets identification.
All large-$R$ jets considered in this analysis are required to have
$\pT > 200$~\GeV and $|\eta| < 1.5$.
Unless otherwise stated, the term “jet” in this paper refers to a large-$R$ jet
that satisfies these selection criteria.

Photons can mimic the signature of emerging jets due to their low associated track activity, and are therefore treated as a potential background. Photon candidates are reconstructed from clustered energy deposits in the electromagnetic calorimeter, either
without a matching ID track or with an associated photon conversion vertex in the ID material. The  \emph{Loose} identification and \emph{Tight} isolation criterion are applied~\cite{EGAM-2018-01}. To mitigate this background, an overlap removal procedure is implemented where large-$R$ jets are discarded if a photon with $\pt > 10$~\GeV is found within $\Delta R = 1.0$ of the jet axis.

Events are required to have a reconstructed primary vertex with at least two
associated tracks~\cite{ATL-PHYS-PUB-2019-015}. The hard scatter (HS) primary vertex is selected as the one with the largest
\(\Sigma \pT^2\), where the sum is over all primary tracks with transverse
momentum \(\pT > \qty{0.5}{\GeV}\) that are associated with the vertex.

Events are selected using one of two unprescaled single jet triggers~\cite{TRIG-2022-01}, both with the same Level 1 requirement that at least 100~\GeV of transverse energy, at the EM scale, is deposited in a region of the calorimeters of size  $0.8\times 0.8$ in $\eta\!-\!\phi$ space~\cite{TRIG-2012-01,TDAQ-2019-01,TRIG-2022-01}.
The first trigger, henceforth referred to as the \textit{high-\pt trigger},
selects events containing a large-$R$ jet and $\pt>460$~\GeV. This
trigger is fully efficient offline for jets with $\pt > 520$~\GeV.
The second, henceforth referred to as the \textit{emerging jet trigger},
selects events containing at least one large-$R$ jet with $\pt > 200$~\GeV, $|\eta| < 1.8$ and a prompt track fraction (PTF) smaller than 0.08.
The PTF is defined as the ratio of the transverse momentum sum of tracks within
$\Delta R=1.2$ of the jet to the jet’s transverse momentum,
\begin{equation}
\textrm{PTF} = \frac{\sum_{\text{trk} \subseteq \Delta R < 1.2} p_{\mathrm{T}}^{\text{trk}}}{p_{\mathrm{T}}^{\mathrm{jet}}}
\label{eq:ptf}
\end{equation}
calculated using the standard track finding algorithm only.
The increased angular distance of $\Delta R=1.2$, compared to the nominal jet
radius of 1.0, accounts for the extended catchment area introduced by the
reclustered small-$R$ jets, which effectively broaden the jet's area.
The tracks entering into the PTF calculation are required to have $\pT > 1$~\GeV,
$|d_0|/\sigma(d_0) < 2.5$, and $\Delta z = |z_{\mathrm{PV}} - z_0| < 10$~mm, where $\sigma(d_0)$ is the uncertainty on the $d_0$ measurement and $z_{\mathrm{PV}}$ is
the position of the HS PV along the $z$-axis. These selections are applied to suppress contributions from pile-up interactions.

A displaced vertex reconstruction algorithm is applied to the combined collection of
tracks from both the primary and large-impact parameter tracking passes~\cite{ATL-PHYS-PUB-2019-013}. To suppress backgrounds from photon conversions and long-lived SM
hadrons, vertices are required to have at least three associated tracks. Additionally, to
reduce contamination from neutral kaons, a vertex mass requirement of $m_{\mathrm{vtx}} >
0.6~\mathrm{\GeV}$ is imposed, where the vertex mass is computed as the invariant mass of
the sum of the four-momenta of the associated tracks, assuming a pion mass hypothesis. A material veto is then applied to
remove vertices whose positions coincide with the location of known detector material. The mapping
of the ID material was performed using Run 2 data~\cite{SUSY-2018-13} and validated with 2022 collision data.
These selection criteria enhance sensitivity to displaced vertices from LLP decays while
effectively suppressing backgrounds from detector interactions and SM
processes. For a given jet, $N_{\mathrm{vtx}}$ is defined as the number of displaced vertices that satisfy
the above criteria and fall within $\Delta R<1.0$ of the jet axis.

%
%
%
%
%
%
%
%
%
%
%
%


%
\section{Analysis strategy}
\label{sec:analysis}

This analysis employs two complementary analysis strategies. The first approach, referred to as the
``cut-based strategy'', identifies candidate EJs based on selections applied to high-level jet
observables, including track-based, vertex-based, and jet substructure-based selections. The second
approach, referred to as the ``ML-based strategy'', uses a per-jet transformer-based ML algorithm trained to differentiate emerging jets from SM jets. While the cut-based method facilitates reinterpretation of the results for alternative theoretical models beyond those
considered in this study, the ML-based strategy maximizes sensitivity for the specific models considered
by this analysis.

For each strategy, two orthogonal event selections are defined based on the dijet invariant mass.
The \lowmjj selection targets lower mediator masses, specifically selecting events with $\mjj < 1~\TeV$.
This region makes use of the emerging jet trigger that has sensitivity to final states with lower \pt jets.
The \highmjj selection targets higher mediator masses, requiring $\mjj > 1~\TeV$. This region is probed
using the high-\pt trigger, which is fully efficient after the offline
selections. All events considered in this analysis are required to have
at least two large-$R$ jets satisfying the definitions given in Section~\ref{sec:objects}.

\subsection{Cut-based strategy}
\label{sec:cutbased_strat}

The cut-based strategy exploits jet-level properties sensitive to both
displacement and the multi-pronged substructure of emerging jets from the decays
of a high multiplicity of long-lived dark pions, enabling effective rejection of
the dominant QCD multijet background.
An initial set of selections, referred to
as preselections, is applied based on these observables. The signal region (SR)
is then defined by applying two additional selections on top of the
preselections. These two variables also define a two-dimensional space referred
to as the \textit{ABCD plane} in which control regions (CRs) are identified. The
CRs are used to estimate the background through a data-driven method known as
the \textit{ABCD method}, described in detail in Section~\ref{sec:background}.

\begin{figure}[h!]
\centering
\begin{subfigure}[b]{0.48\textwidth}
\centering
\includegraphics[width=\textwidth]{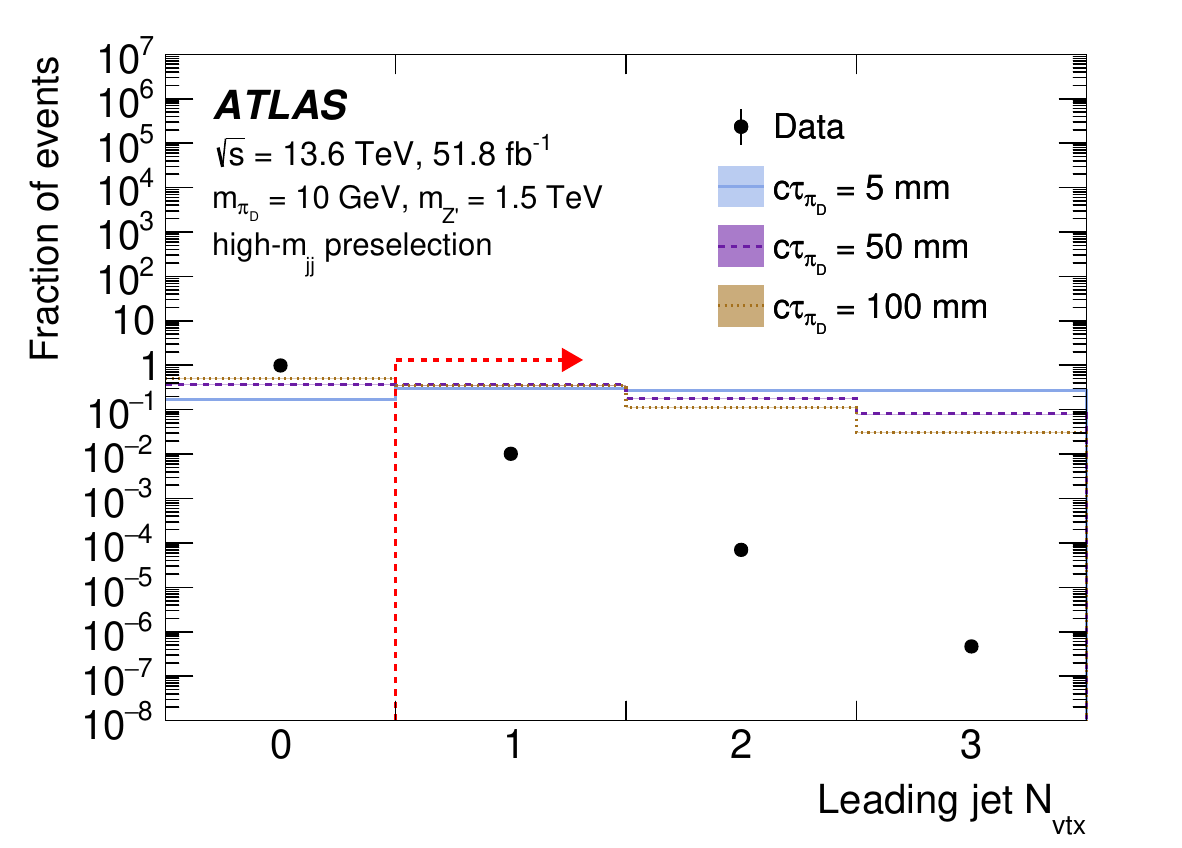}
\label{fig:nvtx:highmjj}
\caption{}
\end{subfigure}
\begin{subfigure}[b]{0.48\textwidth}
\centering
\includegraphics[width=\textwidth]{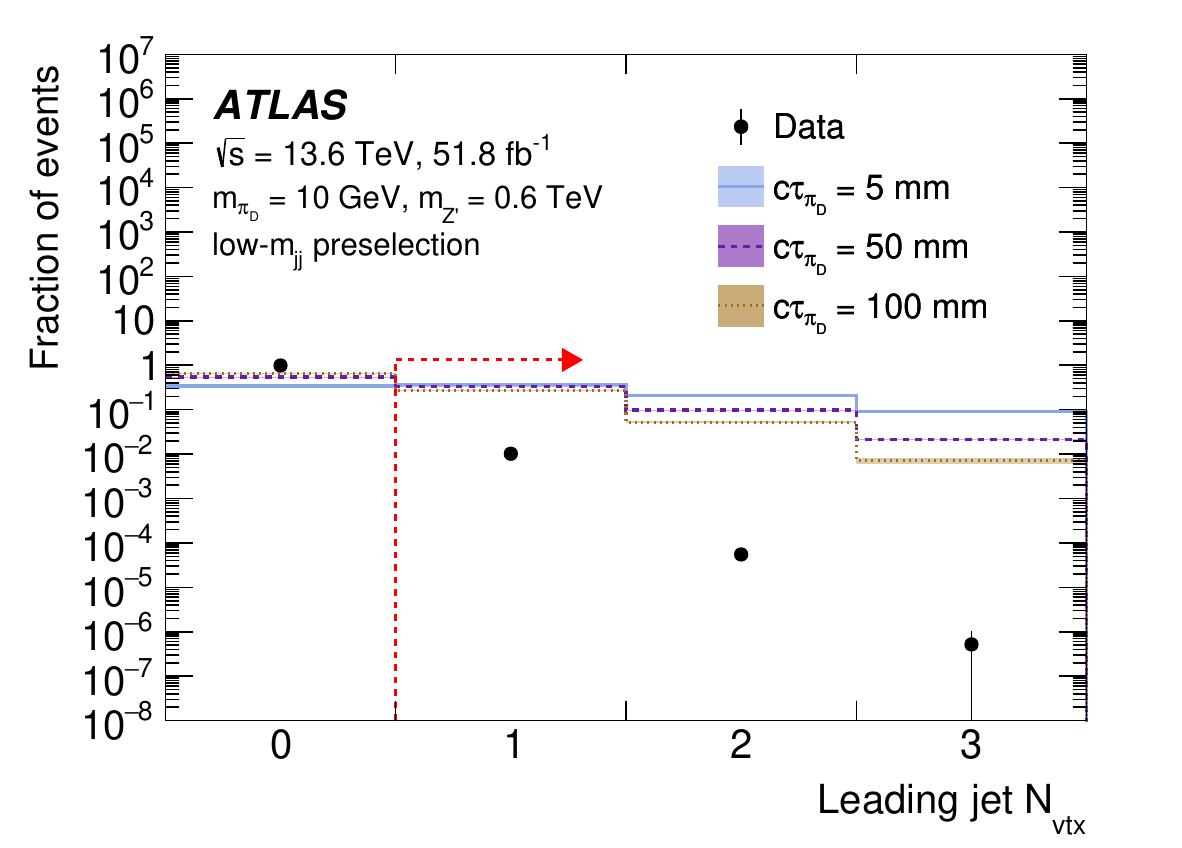}
\label{fig:nvtx:lowmjj}
\caption{}
\end{subfigure}  \\
\caption{Normalised distributions of $N_{\mathrm{vtx}}$ for the leading jet in (a) the \highmjj preselection
and (b) the \lowmjj preselection, shown for $Z'$ signal events (lines) and data (circles). The selection placed on the observable in the
analysis is indicated with a dashed arrow. The rightmost bins include
contributions from overflow events. Similar distributions are observed for the $t$-channel signal and are therefore not shown.}
\label{fig:nvtx}
\end{figure}

In both \highmjj and \lowmjj selections, the leading and
subleading jets are each required to have at least one associated secondary vertex
($N_{\mathrm{vtx}} \geq 1$). The distributions of $N_{\mathrm{vtx}}$ are shown in Figure~\ref{fig:nvtx}.
In both regions, PTF is used to suppress background from SM jets using the same definition as in the trigger (Eq.~\ref{eq:ptf}).
A key challenge in defining the PTF variable is the potential misassignment of
the primary vertex in high-pile-up events.
If a pile-up vertex is incorrectly identified as the HS PV, prompt tracks in the
leading jets may have a large $\Delta z$ relative to the misidentified PV,
resulting in artificially low PTF values and populating the SRs with background
events.
To mitigate this, the difference between the PTF values
with and without the $\Delta z$ requirement is computed:
\begin{equation}
\Delta \PTF = | \PTF - \PTF_{\textrm{no} \ \Delta z \ \textrm{cut}} |
\end{equation}
In both \highmjj and \lowmjj selections, the leading and subleading jets are required to have
$\Delta \PTF < 0.4$ to help reduce background contamination from PV misassignment.
This criterion has minimal impact on signal efficiency, rejecting fewer than 1\% of events.

\subsubsection{\Highmjj region}

In the \highmjj region, events are required to pass the high-\pt trigger. To
ensure full trigger efficiency, the leading jet is required to have $\pt >
520$~\GeV, while the subleading jet must satisfy $\pt > 300$~\GeV to reduce background. Both jets are further required to have $\PTF < 1.0$. To further suppress background contributions, a
selection is applied based on the 2-point energy correlation function
(\ECFtwo)~\cite{Larkoski:2013eya}, defined as:
\begin{equation}
\ECFtwo = \sum_{i<j \in \text{trk}} \pt^{i} \pt^{j} \Delta R_{ij}
\end{equation}
where the sum runs over all pairs of tracks associated with the jet.
\ECFtwo\ quantifies the angular energy dispersion of a jet by measuring pairwise
correlations between the transverse momenta of tracks and their angular
separation. It is particularly sensitive to jets with multi-pronged
substructure, such as those arising from dark quarks. In such cases, the
presence of multiple hard prongs leads to larger \ECFtwo\ values compared to
jets from standard QCD processes, which typically exhibit a more collimated,
single-prong structure. To exploit this feature, a requirement of $\ECFtwo/\pt >
40$~\GeV is imposed on both the leading and subleading jets. This selection
rejects approximately 40\% of background jets while retaining over 90\% signal
efficiency across the benchmark models considered.

As discussed above, the surviving events are mapped onto an ABCD plane defined
by two independent variables, allowing a data-driven background estimation, as
described in Section~\ref{sec:background}. In the \highmjj region, the ABCD
variables are the PTF of the leading and subleading jets. The SR is defined by
requiring both the leading and subleading jet PTF to be less than 0.2.
Distributions of the leading jet PTF are shown in Figure~\ref{fig:jet_vars_a}.
The corresponding subleading jet distributions exhibit similar behaviour and are
therefore omitted. The full set of selections applied in the \highmjj cut-based
region is provided in Table~\ref{tab:mjj_CC_cutlist}. For signal samples with a mediator mass of 1500~\GeV and $m_{\pi_D} = 10$~\GeV, the total signal selection efficiency varies from 0.070\% (0.078\%) in the $s$-channel ($t$-channel) at a lifetime of $c\tau_{\pi_D} = 1000$~mm to 13\% (7\%) at $c\tau_{\pi_D} = 1$~mm  and with a maximum of 29\% (15\%) at $c\tau_{\pi_D} = 5$~mm.
The variation in the signal efficiency  between the $s$-channel and $t$-channel
models arises from differences in the jet $\pt$ spectra and the presence of
additional SM jets in the $t$-channel case, which can be among the
two leading jets and reduce the likelihood that both leading jets are emerging.

\subsubsection{\Lowmjj region}

In the \lowmjj region, events are required to pass the emerging jet trigger. The jets matched to the trigger-level objects are required to have $\pt > 250$~\GeV and PTF $< 0.04$. Since this $\pt$ threshold is
below the trigger efficiency plateau~\cite{TRIG-2022-01}, a $\pt$-dependent trigger efficiency scale factor is applied to each
simulated event.
This scale factor is derived by comparing the emerging jet trigger efficiency in data and samples of simulated $V$+jets events in a dedicated event
selection requiring a single-muon trigger~\cite{TRIG-2022-01}. The efficiency
curves for both data and simulation are fitted using an error function, and the
ratio of the two fitted curves is taken as the trigger scale factor. The scale
factor ranges from 0.72 at $\pt = 250$~\GeV to 0.96 at $\pt = 500$~\GeV.
A systematic uncertainty on the scale factor is derived by varying the fit
parameters within their uncertainties, yielding an uncertainty of approximately
19\% at 250~\GeV and 2\% at 500~\GeV, as discussed in
Section~\ref{sec:systematics}.

\begin{figure}[h!]
\centering
\begin{subfigure}[b]{0.48\textwidth}
\centering
\includegraphics[width=\textwidth]{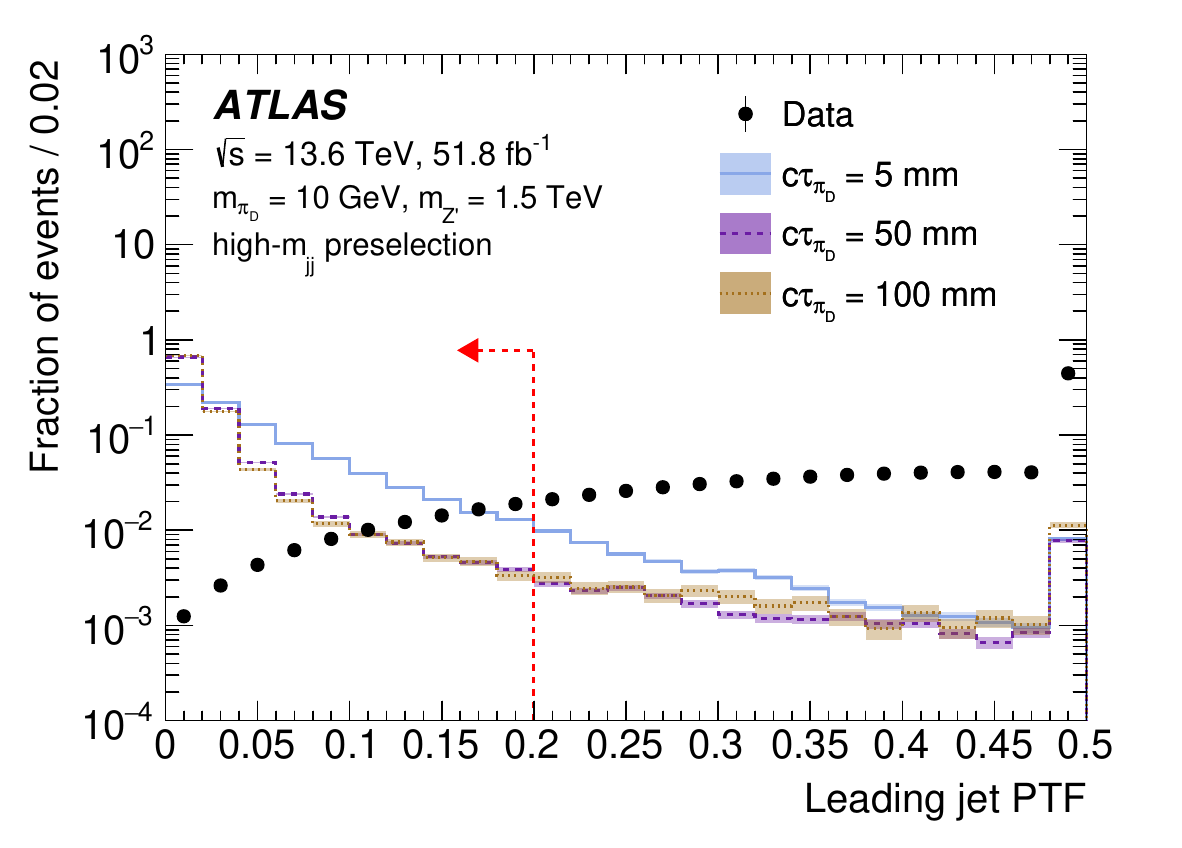}
\caption{}
\label{fig:jet_vars_a}
\end{subfigure}
\begin{subfigure}[b]{0.48\textwidth}
\centering
\includegraphics[width=\textwidth]{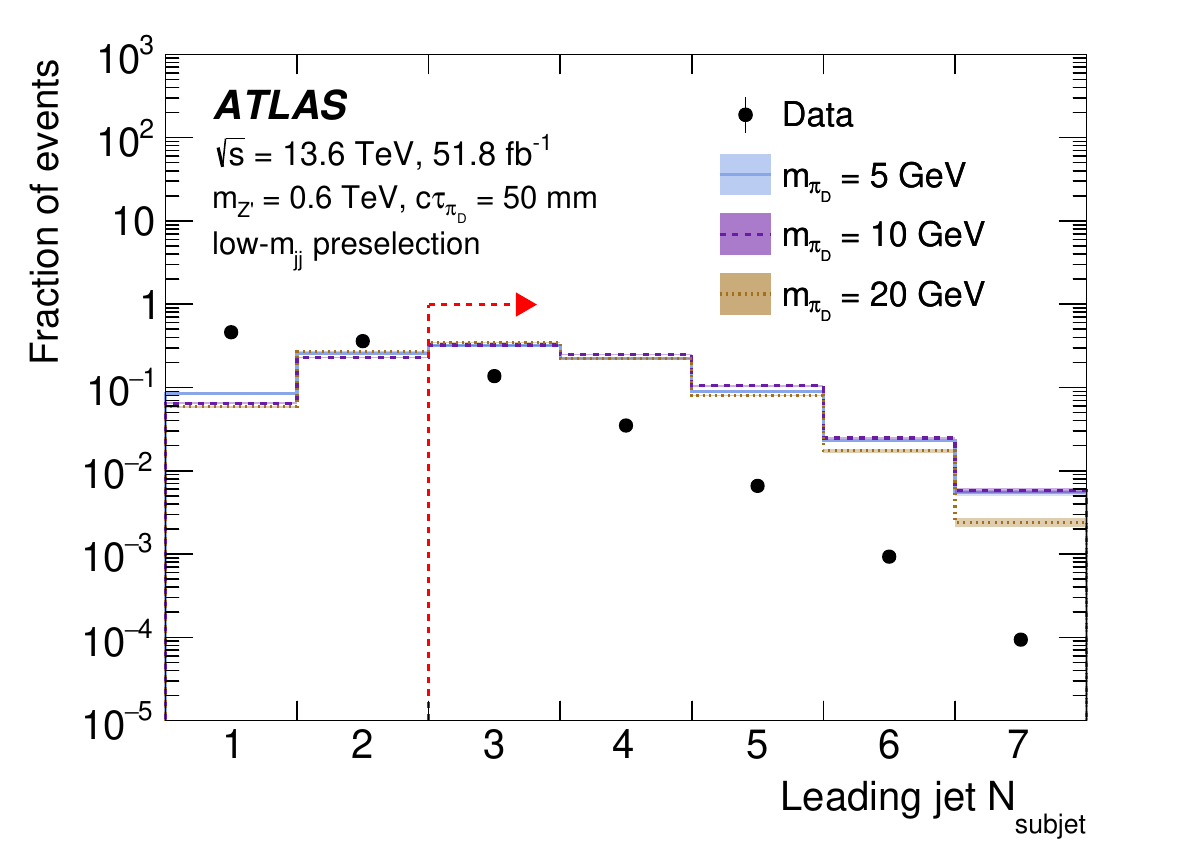}
\caption{}
\label{fig:jet_vars_b}
\end{subfigure}  \\
\caption{Normalised distributions of (a) the leading jet PTF in the
\highmjj preselection and (b) the leading jet number of subjets in the
\lowmjj preselection, shown for $Z'$ signal events (lines) and data (circles). The selection applied to the observable to define the signal
region is indicated by a dashed arrow. The rightmost bins include contributions from overflow events.}
\label{fig:jet_vars}
\end{figure}

Due to the preselection requirement on the PTF of the trigger-matched jet, jets
in the \lowmjj region are biased toward lower PTF values, rendering this
variable unsuitable for defining an ABCD plane. Instead, the \lowmjj ABCD plane
is defined using the number of subjets ($N_{\mathrm{subjet}}$) in the leading
and subleading jets, where ``subjets'' refer to the small-$R$ jets that serve as
constituents of the large-$R$ jets. Emerging jets typically exhibit a higher
number of subjets compared to standard SM jets due to their two-stage parton
shower and hadronisation process. To exploit this property, a minimum of three
subjets per large-$R$ jet is required in the SR selection.
The leading jet $N_{\mathrm{subjet}}$ distributions for signal events and data are shown in Figure~\ref{fig:jet_vars_b}.
Table~\ref{tab:mjj_CC_cutlist} lists the complete set of selections used in the \lowmjj cut-based region.
For signal samples with a mediator mass of 600~\GeV and $m_{\pi_D} = 10$~\GeV, the total signal selection efficiency varies from 0.011\% (0.017\%) in the $s$-channel ($t$-channel) at a lifetime of $c\tau_{\pi_D} = 1000$~mm to a maximum of 8.9\% (3.7\%) at $c\tau_{\pi_D} = 10$~mm and with intermediate eficiencies of 2.6\% (1.5\%) at $c\tau_{\pi_D} = 100$~mm.
In both the \lowmjj and \highmjj selections, the total signal efficiency is
largely independent of $m_{\pi_D}$ for $c\tau_{\pi_D} \lesssim 50$~mm. At longer
lifetimes, however, the increased boost of the $m_{\pi_D} = 5$~\GeV dark pions
causes a larger fraction to decay outside of the tracking fiducial volume, reducing
the efficiency of the analysis selections.

\begin{table}[!htp]
\caption{Selection criteria for the \lowmjj~and \highmjj~regions including the preselections and signal regions for the cut-based strategy. Common criteria are merged across both columns.}
\label{tab:mjj_CC_cutlist}
\centering
\begin{tabular}{lcc}
\toprule{Preselection} & {Low-$m_{jj}$} & {High-$m_{jj}$} \\
\midrule
leading jet $\pt$        & – & $>520$ \GeV \\
subleading jet $\pt$    & – & $>300$ \GeV \\
trigger-matched jet      & $\pt >250$ \GeV, PTF $<0.04$ & – \\
leading and subleading jet PTF  & \multicolumn{2}{c}{$<1.0$} \\
leading and subleading jet $\Delta \PTF$ & \multicolumn{2}{c}{$<0.4$} \\
leading and subleading jet $N_\vtx$ & \multicolumn{2}{c}{$\geq 1$} \\
leading and subleading jet $\ECFtwo/\pt$ & – & $>40$ \GeV \\
$m_{jj}$ & $<1$ \TeV & $>1$ \TeV \\
number of jets & \multicolumn{2}{c}{$\geq 2$} \\
\midrule
\multicolumn{3}{l}{Signal Region} \\
\midrule
leading and subleading jet $N_{\text{subjet}}$ & $\geq 3$ & – \\
leading and subleading jet PTF & – & $<0.2$ \\
\bottomrule
\end{tabular}
\end{table}

\subsection{ML-based strategy}
\label{sec:ML_strategy}

The ML-based strategy uses a transformer jet tagging algorithm based on the ATLAS GN2 flavour
tagging algorithm~\cite{FTAG-2023-05}.
The input to the model consists of jet
features concatenated with the feature vectors of up to 200 associated tracks.
These feature vectors are processed by a per-track initialization network,
which includes a hidden layer and an output layer of size 256, increasing each
tracks' embedding dimension. The embedded track representations are then passed
through a five-layer transformer encoder with eight attention heads to produce
conditional track representations based on the other tracks in the jet. The
resulting track embeddings are projected to a dimension of 128, and a global
jet representation is computed using attention pooling~\cite{Li2017}.

The complete list of jet and track features is provided in Table~\ref{tab:GNN_inputs}. The
only jet feature included is pseudorapidity, which helps to account for tracking efficiency
effects related to detector acceptance. Jet \pt, an input in the GN2 algorithm, is
excluded as it was found to provide no improvement in classification performance while
increasing the tagger's dependence on jet momentum.
\begin{table}[h!]
\caption{List of the track and jet variables used in the transformer-based emerging jet tagging algorithm. Each track in the jet has its own input feature vector with this information. The first entry is a jet level input, which is common to all of the tracks. ``Shared hits`` refer to hits that are common between
multiple reconstructed tracks.}
\label{tab:GNN_inputs}
\centering
\renewcommand{\arraystretch}{1.2}
\begin{tabular}{c|c}
\toprule
Input & Description \\
\hline
Jet $\eta$ & Jet pseudorapidity\\
\hline
$d_0$ & Track closest distance to PV in transverse plane\\
\hline
$z_0 \text{sin}(\theta)$ & Track closest distance to PV in longitudinal plane \\
\hline
$\Delta\phi$ & Azimuthal angle of the track, relative to the jet $\phi$\\
\hline
$\Delta\eta$ & Track pseudorapidity, relative to jet $\eta$\\
\hline
$q/p$ & Track charge over momentum\\
\hline
$\sigma(\phi)$ & Uncertainty in track $\phi$\\
\hline
$\sigma(\theta)$ & Uncertainty in track $\theta$\\
\hline
$\sigma(q/p)$ & Uncertainty in track $q/p$\\
\hline
$d_{0}/\sigma(d_0)$ & signed $d_{0}$ significance\\
\hline
$z_{0}/\sigma(z_0)$ & signed $z_{0}$ significance\\
\hline
$N_{\textrm{PIX hits}}$ & Number of Pixel hits per track\\
\hline
$N_{\textrm{SCT hits}}$ & Number of SCT hits per track\\
\hline
$N_{\textrm{IBL hits}}$ & Number of innermost pixel layer hits\\
\hline
$N_{\textrm{PIX shared}}$ & Number of Pixel shared hits\\
\hline
$N_{\textrm{SCT shared}}$ & Number of SCT shared hits\\
\bottomrule
\end{tabular}
\end{table}

The algorithm optimises three tasks simultaneously: jet classification, track
origin classification, and track-pair compatibility. The primary jet
classification network, which operates on the pooled jet representation, outputs
the probability that a given jet is an emerging jet, $p_{\mathrm{EJ}}$, which is
referred to as the jet classification score.
The track origin classification network assigns each track to one of four
categories based on its associated truth-matched particle~\cite{IDTR-2021-03}:
prompt (originating from the hard-scatter event but not from a dark pion decay),
pile-up (originating from a pile-up vertex), fake (no valid truth match), and
displaced (originating from a dark pion decay).
The track-pair
compatibility network predicts whether two tracks originate from the same vertex
using a binary classifier. Each task-specific network consists of three hidden
layers with sizes 128, 64, and 32, using Rectified Linear Unit (ReLU) activation
functions throughout~\cite{Agarap:2018uiz}. A combined cross-entropy loss function with tunable
weights ensures simultaneous optimisation of all three tasks. The output of the
auxiliary tasks is not directly used in the selection of emerging jet candidates
in the analysis, but it plays a crucial role in improving the convergence of the
jet classification task by contributing to the overall loss function.

This is the first application of the GN2 architecture to train a dedicated
tagger for a beyond-the-Standard-Model signature, demonstrating the flexibility
of the architecture beyond its original use in jet flavour tagging.  It is also
the first ML tagger in ATLAS to incorporate information from the secondary
large-impact parameter tracking pass, and the first use of ML to identify
displaced secondary vertices in an ATLAS LLP search. This approach is broadly
applicable and has the potential to enhance sensitivity across a wide range of
LLP signatures within ATLAS.

The model is trained on 12 million jets, equally sampled from simulated SM multijet events and \( pp
\rightarrow Z' \rightarrow q_D \bar{q}_D \) events taken from samples with $m_{Z'} \in \{600, 1500, 3000\}~\mathrm{\GeV}$ and $c\tau_{\pi_D} \in \{5,50\}~\mathrm{mm}$. To prevent evaluation bias, the dataset is split into two
disjoint folds (\( f_0 \), \( f_1 \)), each containing six million jets. Two separate instances of the tagger
are trained, one per fold, with 5.5 million jets for training and 500,000 for validation. During
evaluation, jets used to train \( f_0 \) are tested with \( f_1 \), and vice versa. The two folds yield
statistically consistent performance.

\begin{figure}[h!]
\centering
\includegraphics[width=0.85\textwidth]{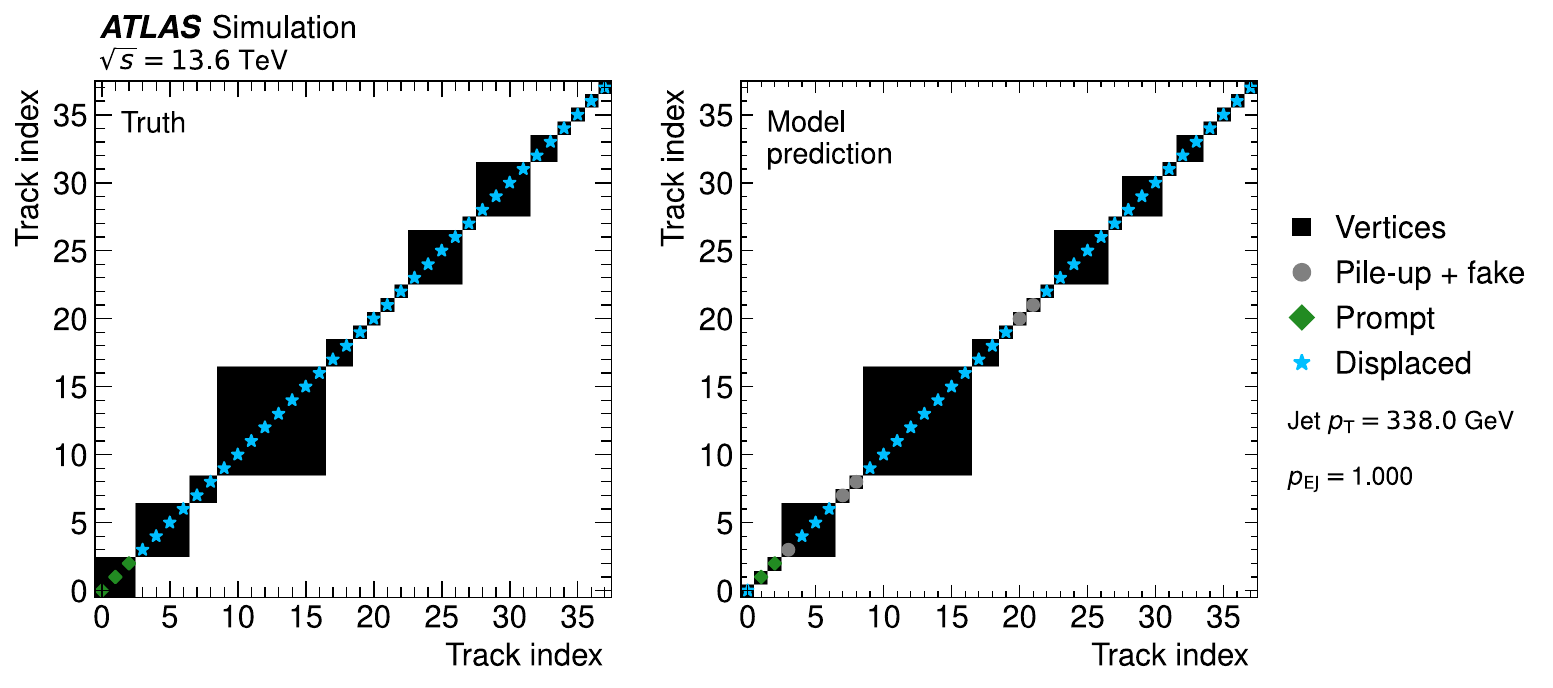}
\caption{Example of the auxiliary task performance on a single jet with $p_{\mathrm{EJ}} = 1.000$ from
a sample of simulated signal events with $m_{Z'} = 1500$~\GeV, $m_{\pi_D} = 10$~\GeV, and $c\tau_{\pi_D} = 50$~mm. The reconstructed tracks in the jet
are shown along the diagonal and are displayed using different markers based on their origin, with
displaced tracks shown as stars, prompt tracks shown as diamonds, and both pile-up
and fake tracks shown as circles. Groups of tracks that form a common vertex
are shown as filled squares. The left diagram shows the true track and
vertex labels, and the right diagram shows the model predictions. Track
indices are ordered such that tracks originating from the same truth vertex
appear consecutively.}
\label{fig:jet-view}
\end{figure}

Figure~\ref{fig:jet-view} shows a matrix representation comparing the true and
predicted classifications of tracks and vertex groupings for an example signal
jet with a classification score of $p_{\mathrm{EJ}} = 1.000$. The tracks are
displayed along the diagonal and are distinguished by marker style according to
their origin: pile-up/fake, prompt, or displaced. Vertex groupings are marked by
black areas and are formed using the track pair-wise compatibility scores via a
union-find algorithm~\cite{Kozen1992}. The auxiliary tasks exhibit high
performance on this example jet, with 85\% of tracks correctly labeled and 79\%
successfully assigned to their respective vertices. Across the full test
dataset, displaced tracks are correctly labeled with an average accuracy of
91\%, and 80\% are successfully grouped into their corresponding vertices.

Similar to the cut-based strategy, events are categorised into \highmjj ($\mjj >
1$~\TeV) and \lowmjj ($\mjj < 1$~\TeV) regions. In the \highmjj region, events are
required to pass the high-\pt trigger, with the leading jet satisfying $\pt >
520$~\GeV and the subleading jet having $\pt > 300$~\GeV. In the \lowmjj region, events
must pass the emerging jet trigger, and the jet that is matched to the trigger-level object is required
to have $\pt > 300$~\GeV and PTF $< 0.04$. The \pt threshold in the ML-based
strategy is higher than in the cut-based approach, as the background rejection
of the jet classification task is observed to degrade at low \pt, as discussed
in Section~\ref{sec:background}. These requirements define the preselection for the ML-based strategy. No selection is applied on $N_{\mathrm{vtx}}$
or $\Delta \PTF$.

\begin{figure}[h!]
\centering
\includegraphics[width=0.5\textwidth]{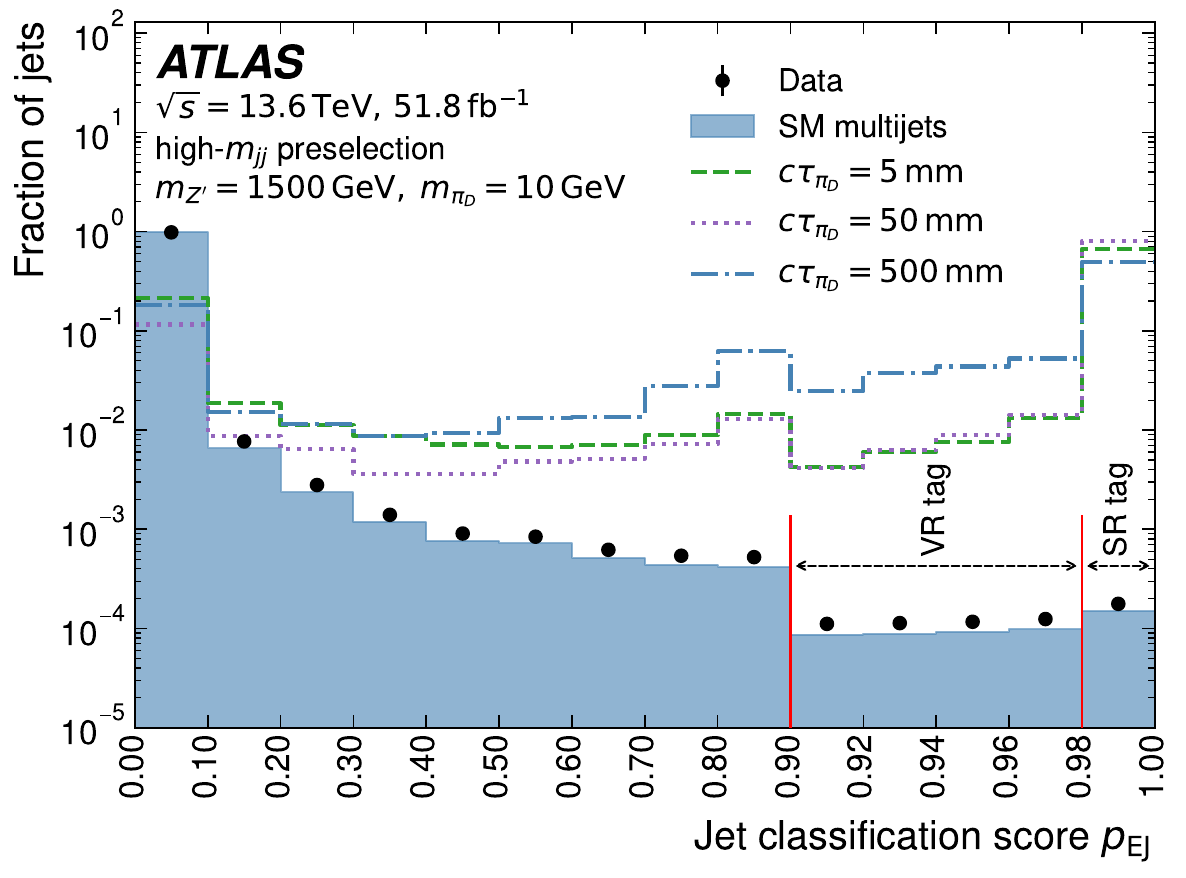}
\caption{Normalised distribution of the jet classification score for all jets in data,
simulated SM multijets background, and selected signal samples with $m_{Z'} = 1500$~\GeV and $m_{\pi_D} = 10$~\GeV
passing the ML-based \highmjj preselection. Vertical lines indicate the
selections placed on the classification score used to define a validation tag
and a signal tag. No uncertainties are shown on the background MC, which is intended for qualitative comparison only.}
\label{fig:gnn_score}
\end{figure}

The distribution of the jet classification score is shown in Figure~\ref{fig:gnn_score}.
The probability \( p_{\mathrm{EJ}} \) is used to tag emerging jet candidates with a
threshold of 0.98. Jets with $0.90 < p_{\mathrm{EJ}} < 0.98$ are used
to define a validation region (VR) tag, as described in Section~\ref{sec:background}. For both \lowmjj and \highmjj regions, a SR is defined
by requiring \( n_{\mathrm{tag}} \geq 2 \), where \( n_{\mathrm{tag}} \) is the
number of tagged jets in the event, and a CR is defined by requiring \(
n_{\mathrm{tag}} < 2 \). The threshold of 0.98 was chosen to optimise background
rejection while ensuring that the level of signal contamination in the control region
remained below 10\%, assuming signal cross sections to which the analysis is
expected to be sensitive. This threshold yields a 97.5\% EJ tagging efficiency
and a 0.04\% background tagging efficiency (corresponding to a rejection
factor of 2600) when evaluated on the testing dataset.

The complete set of selections for the ML-based analysis are listed in
Table~\ref{tab:gnn_selections}. For samples with $m_{Z'} = 1500$~\GeV and
$m_{\pi_D} = 10$~\GeV, the total signal selection efficiency considering both
\lowmjj and \highmjj regions ranges from 28\% at $c\tau_{\pi_D} = 1$~mm to 0.8\%
at $c\tau_{\pi_D} = 1000$~mm, reaching a maximum of 35\% at $c\tau_{\pi_D} =
10$~mm. In the case of the $t$-channel scalar mediator scenario with $m_{\Phi} =
1500$~\GeV, the efficiency spans from 53\% to 2.5\% over the same lifetime range,
with a maximum of 65\% observed at $c\tau_{\pi_D} = 10$~mm.
At high $c\tau_{\pi_D}$, the efficiency declines primarily because many dark
pions decay outside the tracking fiducial region, preventing the reconstruction
of their charged decay products. Additionally, for $c\tau_{\pi_D} = 1000$~mm,
some dark pion decays will occur inside or beyond the calorimeter, thereby
reducing the visible energy available for jet reconstruction.

\begin{table}[!htp]
\caption{Selection criteria for the \lowmjj~and \highmjj~regions in the ML-based strategy. Common criteria are merged across both columns.}
\label{tab:gnn_selections}
\centering
\begin{tabular}{lcc}
\toprule{Variable} & {Low-$m_{jj}$} & {High-$m_{jj}$} \\
\midrule
leading jet $\pt$        & $>300$ \GeV & $>520$ \GeV \\
trigger-matched jet      & $\pt >300$ \GeV \& PTF $<0.04$ & - \\
sub-leading jet $\pt$     & $>200$ \GeV & $>300$ \GeV \\
$m_{jj}$              & $<1$ \TeV & $>1$ \TeV \\
number of jets   & \multicolumn{2}{c}{$\geq 2$} \\ %
number of tagged jets ($p_{\mathrm{EJ}}>0.98$) & \multicolumn{2}{c}{CR: 0, 1; SR: $\geq 2$} \\ %
\bottomrule
\end{tabular}

\end{table}

%
%
%


%
\section{Background estimation}
\label{sec:background}

In both the cut-based and ML-based analyses, the dominant background arises from
SM multijet production. Due to the difficulties in accurately modelling this
background with simulation, data-driven techniques are used for background
estimation in both strategies.

\subsection{Cut-based strategy}
\label{sec:bkg_cutbased}
In the cut-based strategy, the data-driven ABCD method is employed to estimate the
contribution from SM background in each of the two selections.
The method employs a 2D parameter space divided into four regions—$A$, $B$, $C$, and $D$—by applying selection criteria on each of the two variables. If the background distributions of these variables are independent, the expected background yield in one region is mathematically related to those in the three other regions. Specifically, when the background primarily populates regions $B$, $C$, and $D$, while the signal is concentrated in region $A$, the background yield in $A$ is estimated as:
\begin{equation}
\label{eq:abcd_pred}
N_A = \frac{N_B \times N_C}{N_D}
\end{equation}
This formula is thus employed in both \lowmjj and \highmjj selections to predict the background in their respective SRs, $A$. When performing the final statistical analysis, any signal contamination in the background regions is
accounted for using a simultaneous fit, as described in Section~\ref{sec:results}.

\begin{figure}[h!]
\centering
\begin{subfigure}[b]{0.485\textwidth}
\centering
\includegraphics[width=\textwidth]{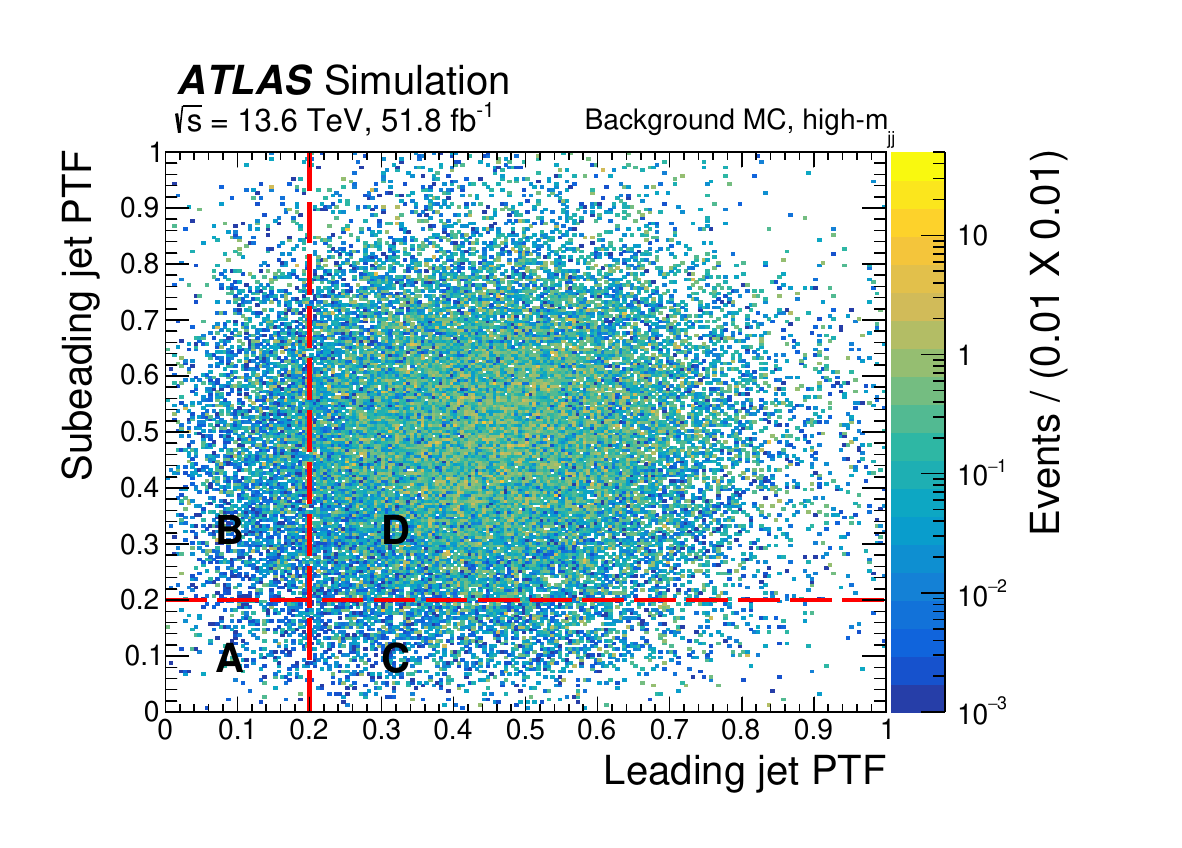}
\caption{}
\label{fig:ABCD:bkg:highmjj}
\end{subfigure}
\begin{subfigure}[b]{0.485\textwidth}
\centering
\includegraphics[width=\textwidth]{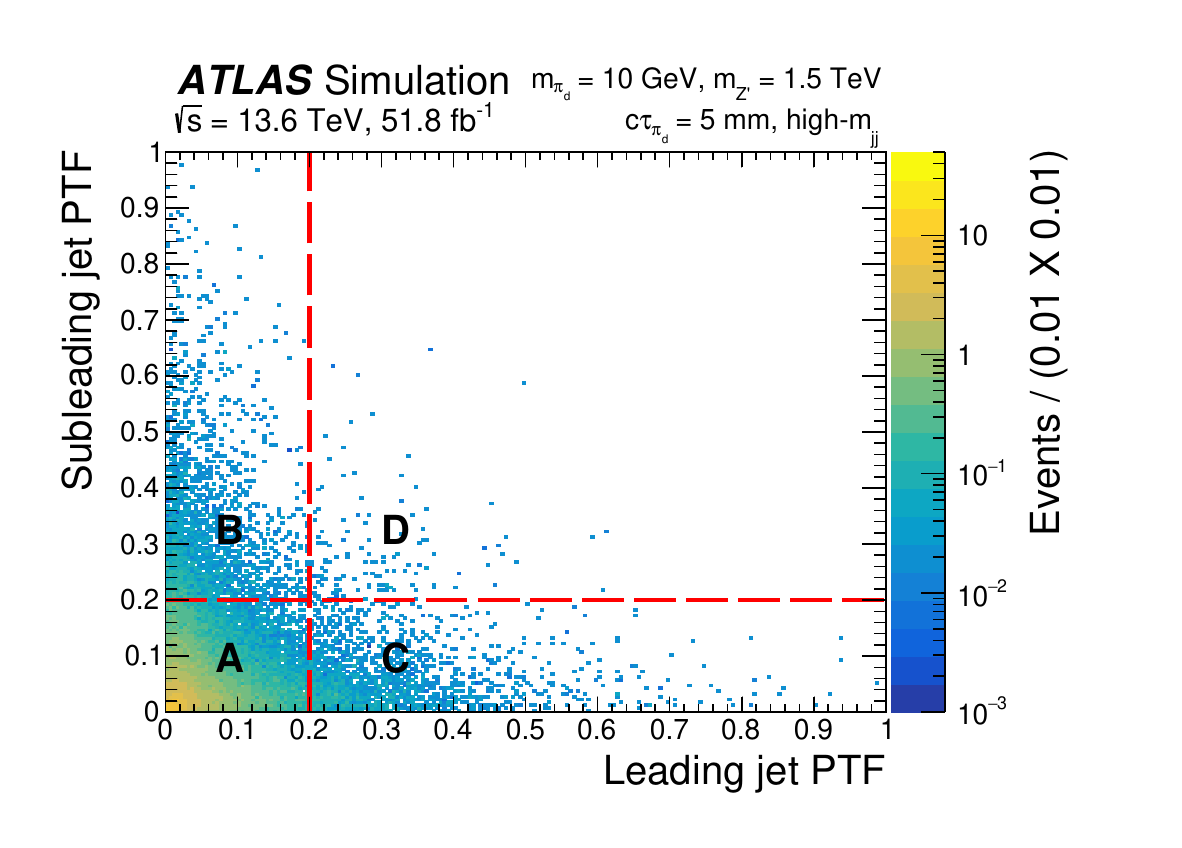}
\caption{}
\label{fig:ABCD:sig:highmjj}
\end{subfigure}  \\
\begin{subfigure}[b]{0.485\textwidth}
\centering
\includegraphics[width=\textwidth]{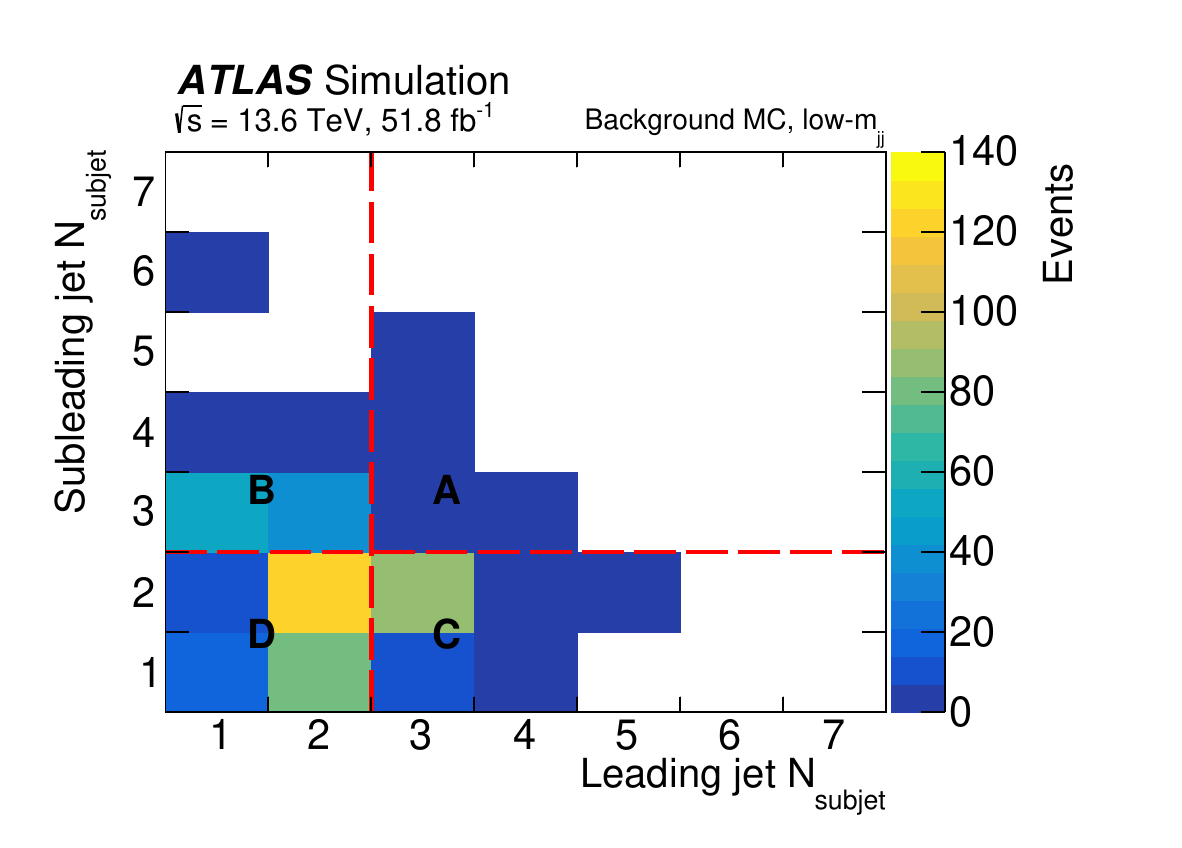}
\caption{}
\label{fig:ABCD:bkg:lowmjj}
\end{subfigure}
\begin{subfigure}[b]{0.485\textwidth}
\centering
\includegraphics[width=\textwidth]{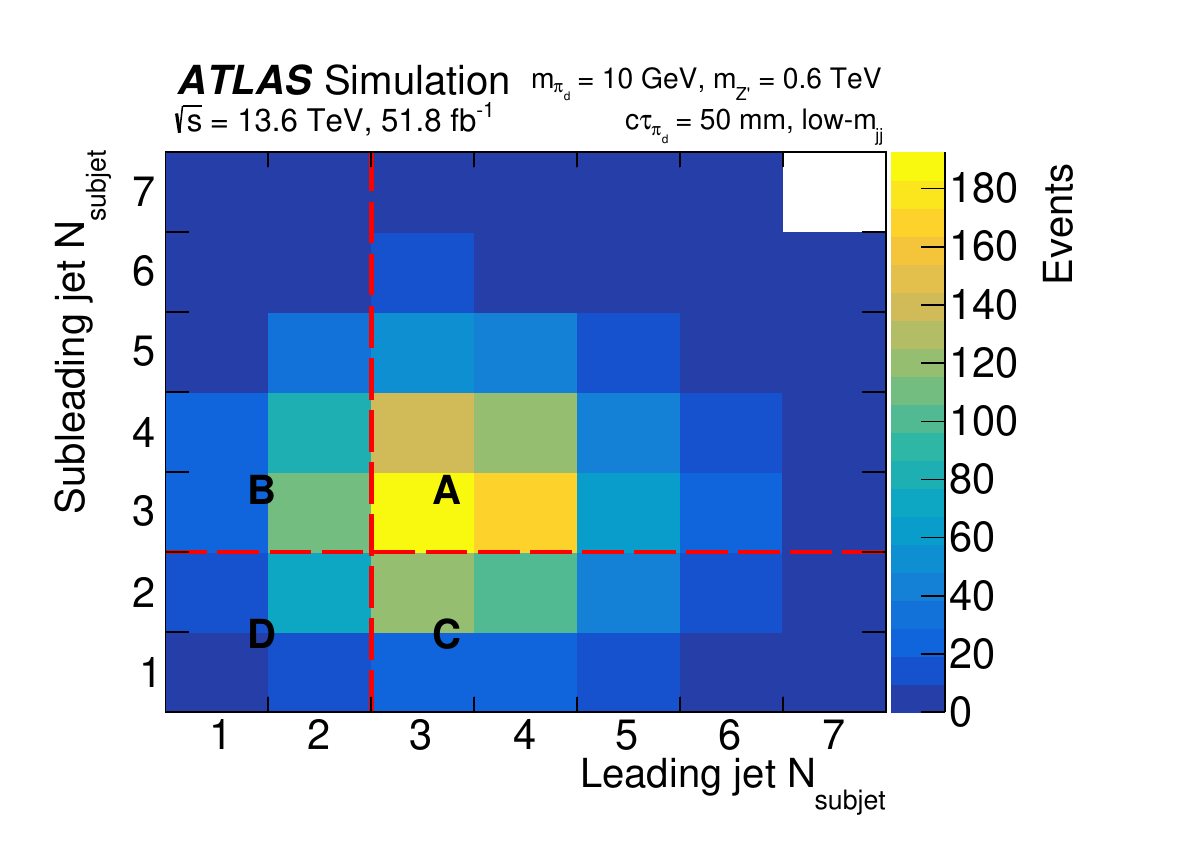}
\caption{}
\label{fig:ABCD:sig:lowmjj}
\end{subfigure}
\caption{The distributions of (a,c) simulated background events (b,d) and signal events in the (a,b) \highmjj region and (c,d) \lowmjj region. The parameters of the simulated signal shown in (b) corresponds to $m_{Z'} = 1.5~\TeV$, $\mpiD = 10~\GeV, \ctauD = 5~\text{mm}$ and the signal shown in (d) corresponds to $m_{Z'} = 0.6~\TeV, \mpiD = 10$~\GeV and $\ctauD = 50$~mm.}
\label{fig:ABCD}
\end{figure}

As described in Section~\ref{sec:cutbased_strat}, the ABCD plane is constructed with the $N_{\mathrm{subjet}}$ variable for the leading and subleading jets in the \lowmjj selection and the PTF variable in the \highmjj selection. The distribution of events in the ABCD planes as well as the SR and CR are shown in Figure~\ref{fig:ABCD} for both the \highmjj and \lowmjj selections.

\begin{figure}[h!]
\centering
\includegraphics[width=\textwidth]{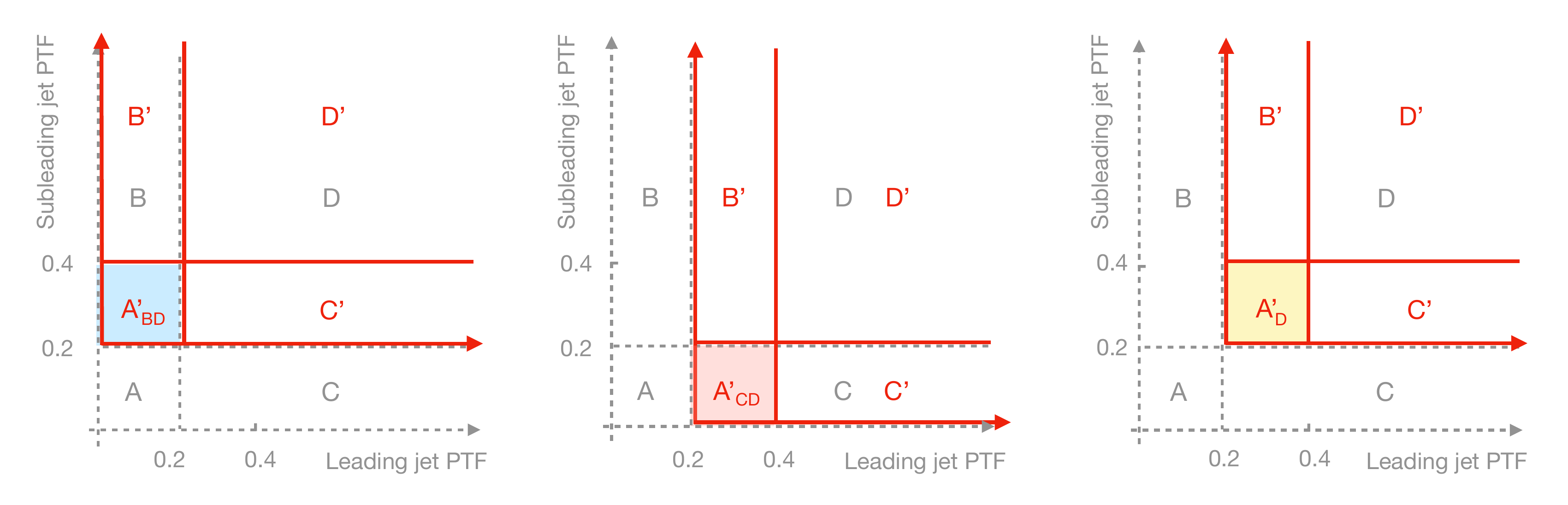}
\caption{Schematic definitions of the $A'$ control regions within the nominal ABCD plane of the \highmjj selection.}
\label{fig:primed_ABCD}
\end{figure}

To validate the background estimation method, control regions from the standard ABCD plane are used in both the \lowmjj and \highmjj event selections. In the \highmjj case, three modified ABCD planes are defined to probe regions closer to the SR while still being dominated by background.
These planes are constructed within merged ABCD control regions—specifically, regions BD, CD, and D—and are used to define new signal-like regions, denoted as $A'_{BD}$, $A'_{CD}$, and $A'_{D}$.
The region $A'_{CD}$ is defined by requiring a subleading jet PTF below 0.2 and a leading jet PTF in the range $0.2 < \mathrm{PTF}_{\mathrm{lead}} < 0.4$.
In $A'_{BD}$ events are selected with subleading jet PTF in the range $0.2 < \mathrm{PTF}_{\mathrm{sublead}} < 0.4$, while the leading jet PTF remains below 0.2.  Finally, $A'_{D}$ is defined by selecting events where both the leading and subleading jet PTFs lie within $0.2 < \mathrm{PTF} < 0.4$. A sketch of the signal-like $A'$ regions is shown in Figure~\ref{fig:primed_ABCD}.

These regions are designed to be the closest approximations to the nominal SR
within their respective modified ABCD planes, thus serving as critical areas for
validating the background estimation. The ABCD method is applied in each
modified plane to estimate the event yield in the corresponding $A'$ regions.
The predicted and observed yields are summarised in Table~\ref{tab:ap_results}.
In each CR, the significance of the deviation is quantified by the absolute difference between the predicted and observed yields, normalised by the quadrature sum of the statistical uncertainty on the prediction and the Poisson uncertainty on the observed yield.
To estimate a representative potential nonclosure significance for the SR,
the average of these normalised absolute differences across the three $A'$ regions is
computed, yielding a value of 1.0. An additional systematic uncertainty is therefore
assigned to the SR background prediction, taken as $1.0 \times
\sigma_\text{stat}(N_{\text{pred},A})$.

\begin{table}[t!]
\centering
\caption{Predicted and observed yields in the $A'_{BD}$, $A'_{CD}$, and $A'_{D}$ regions of the \highmjj selection, which are closest to the nominal signal region. The predictions are obtained from the ABCD relation in Equation \ref{eq:abcd_pred}.}
\label{tab:ap_results}
\begin{tabular}{l|ccc}
\toprule
& $A'_{BD}$ & $A'_{CD}$ & $A'_{D}$ \\
\midrule
Predicted & $43 \pm 5~\mathrm{(stat.)}$ & $26 \pm 4~\mathrm{(stat.)}$ & $160 \pm 14~\mathrm{(stat.)}$ \\
Observed  & $49$ & $34$ & $182$ \\
\bottomrule
\end{tabular}
\end{table}

Similarly, in the \lowmjj region, control regions are merged to define a BD and CD portion of the nominal ABCD plane to test the background estimation strategy. However, due to the
discontinuity of the ABCD plane and the limited statistics, constructing a statistically robust test with $A'$ regions closer to the nominal SR is challenging. To mitigate this, the ensemble of $A'$ regions is rather
defined by systematically varying the selection thresholds on $N_\mathrm{subjet}$, resulting in eight different $A'$ regions and reducing the impact of
individual fluctuations on the nonclosure uncertainty. Averaging over these eight tests, the mean absolute discrepancy
is found to be 1.0 times the statistical uncertainty of the difference in prediction. Consequently, a
nonclosure uncertainty of $1.0 \times
\sigma_\text{stat}(N_{\text{pred},A})$ is assigned to the background estimate in the \lowmjj SR.

Finally, in both the \highmjj and \lowmjj regions, additional validations are performed
using an ML-inverted VR, where the jet classification scores of the
leading and subleading jets in the ABCD plane are required to be less than 0.95. It is
verified in simulation that inverting this requirement removes the vast majority signal events while
leaving the background nearly unchanged. Consequently, this defines a signal-free plane with
a background composition closely matching that of the nominal region, enabling a robust
validation of the ABCD method on this background-enriched sample, including in a region $A'$
that closely resembles region $A$. In both the \highmjj and \lowmjj regions, the observed
yields in the ML-inverted $A'$ region are found to be in agreement with the ABCD predictions
within uncertainties, as shown in Table~\ref{tab:gnn_inverted_results}. The ABCD background
estimates were further validated using simulated background samples and were found to be
consistent within statistical uncertainties.
\begin{table}[h]
\centering
\caption{Predicted and observed yields in the $A'$ region of the ML-inverted VR in the \highmjj and \lowmjj selections. The predictions are obtained from the ABCD relation in Equation \ref{eq:abcd_pred} and the systematic uncertainties are the nonclosure uncertainties described in the text.}
\label{tab:gnn_inverted_results}
\begin{tabular}{l|cc}
\toprule
&  \lowmjj VR & \highmjj VR \\ \midrule
Predicted  & $12.2 \pm 3.9~\mathrm{(stat.)} \pm 3.9~\mathrm{(syst.)} $   & $7.5 \pm 1.1~\mathrm{(stat.)} \pm 1.1~\mathrm{(syst.)} $  \\
Observed   & $7$           & $8$          \\
\bottomrule
\end{tabular}
\end{table}

\subsection{ML-based strategy}
\label{sec:bkg_gnn}

In the ML-based strategy, the jet classification score is correlated with several jet-level features, including the $\pt$ and
PTF of  the jet, as well as the jet flavour. These correlations therefore lead to
correlations between the tagger scores of individual jets in each background event.
Hence, the ABCD method is not well-suited for predicting the expected background in the
GNN-based SRs. Instead, a different data-driven  background estimation strategy is used,
which relies on the determination of the probability that a given background jet will be
misclassified as signal, a
quantity henceforth referred to as the \textit{mistag rate}. The mistag rate is measured directly in
data using the $n_{\mathrm{tag}} < 2$ CR by computing the ratio of tagged jets to all jets in this
region.

After calculating the mistag rate, the probability of tagging exactly one or exactly zero jets in an event can be calculated from
\begin{align}
P(\text{1 tag}|\text{event}) &= \sum_{i = 1}^{n_{\mathrm{jet}}} P(\text{tag}|j_i)
\times \prod_{k \neq i}^{n_{\mathrm{jet}}} \left(1 - P(\text{tag}|j_k) \right) \\
P(\text{0 tag}|\text{event}) &= \prod_{i=1}^{n_{\mathrm{jet}}} \left(1 -  P(\text{tag}|j_i) \right) \label{eq:prob1DV}
\end{align}
where $n_{\mathrm{jet}}$ is the number of jets in the event and $P(\text{tag}|j_i)$
is the mistag rate of jet $i$ in an event
($i=1,\ldots,n_{\mathrm{jet}}$). The probability of obtaining two or more tags is
given by the  complement of $P(\text{1 tag}|\text{event})$ and $P(\text{0
tag}|\text{event})$:
\begin{equation}
\label{eq:prob2DV}
P(\geq\text{2 tag}|\text{event}) = 1 -P(\text{1 tag}|\text{event}) - P(\text{0 tag}|\text{event})
\end{equation}

To get a prediction on the number of events in the one and two tag regions, the probabilities
$P(\text{1 tag}|\text{event})$ and $P(\geq\text{2 tag}|\text{event})$ calculated from each preselected event
are used as per-event weights, with the sum of weights comprising the final background
prediction.

Mistag rates are calculated differentially based on jet-level observables. As previously mentioned, the tagger score is correlated with several jet observables, which can also be correlated between individual jets in an event. The dominant correlations include jet $\pt$, jet PTF, the number of heavy-flavour decays, the number of reconstructed secondary vertices, and the total number of tracks associated with the jet. For example, at high $\pt$, SM jets have a large number of high-momentum tracks with low \( |d_0| \), making it easier to distinguish between signal and background jets. As the jet $\pt$ decreases, the number of prompt tracks decreases, leading to an increased mistag rate. Conversely, if a background jet contains more heavy-flavour hadrons, the displaced vertices that mimic dark pion decays make it harder to discriminate between signal and background, thereby increasing the mistag rate.

\begin{figure}[h!]
\centering
\begin{subfigure}[b]{0.425\textwidth}
\centering
\includegraphics[width=\textwidth]{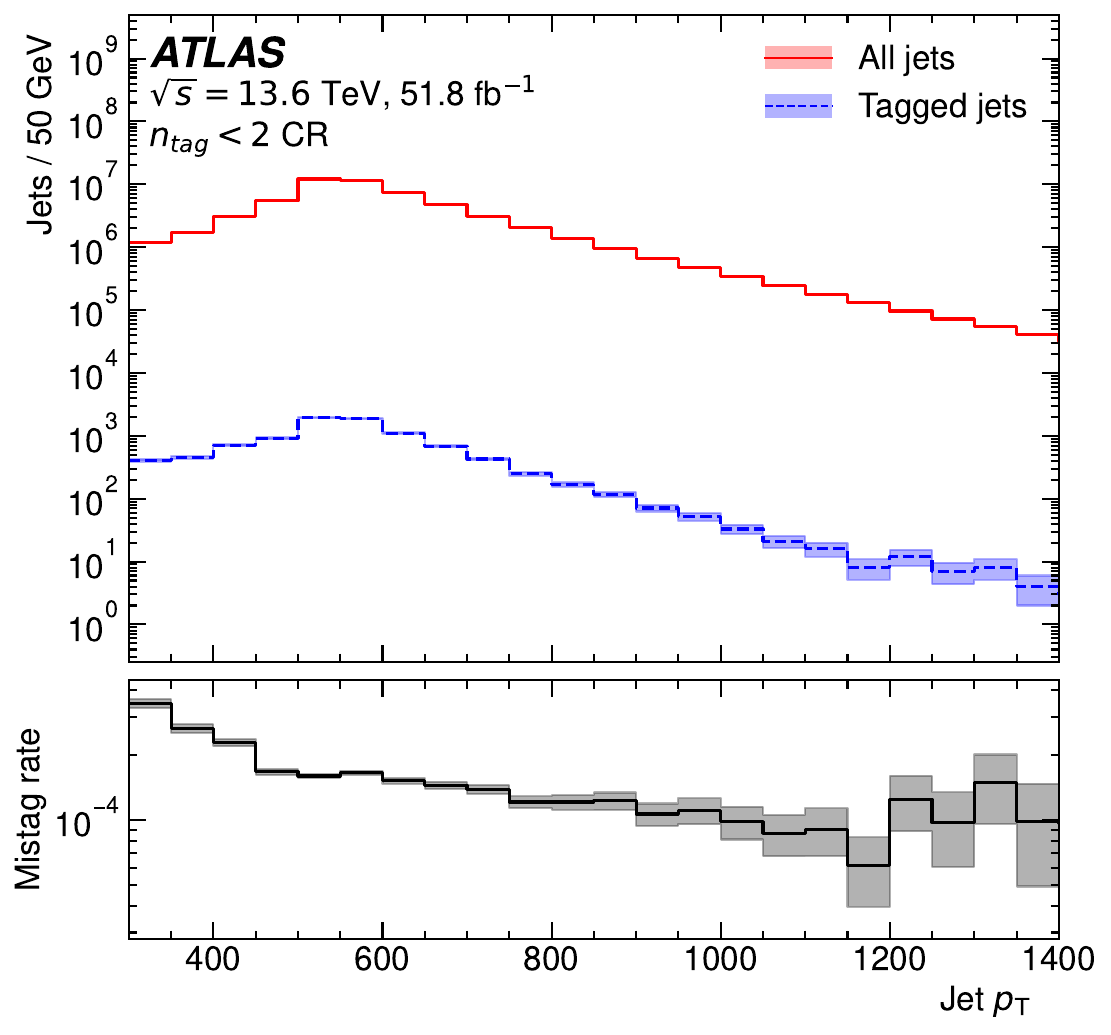}
\caption{}
\label{fig:gnn_mistag_correlation:pt}
\end{subfigure}
\begin{subfigure}[b]{0.425\textwidth}
\centering
\includegraphics[width=\textwidth]{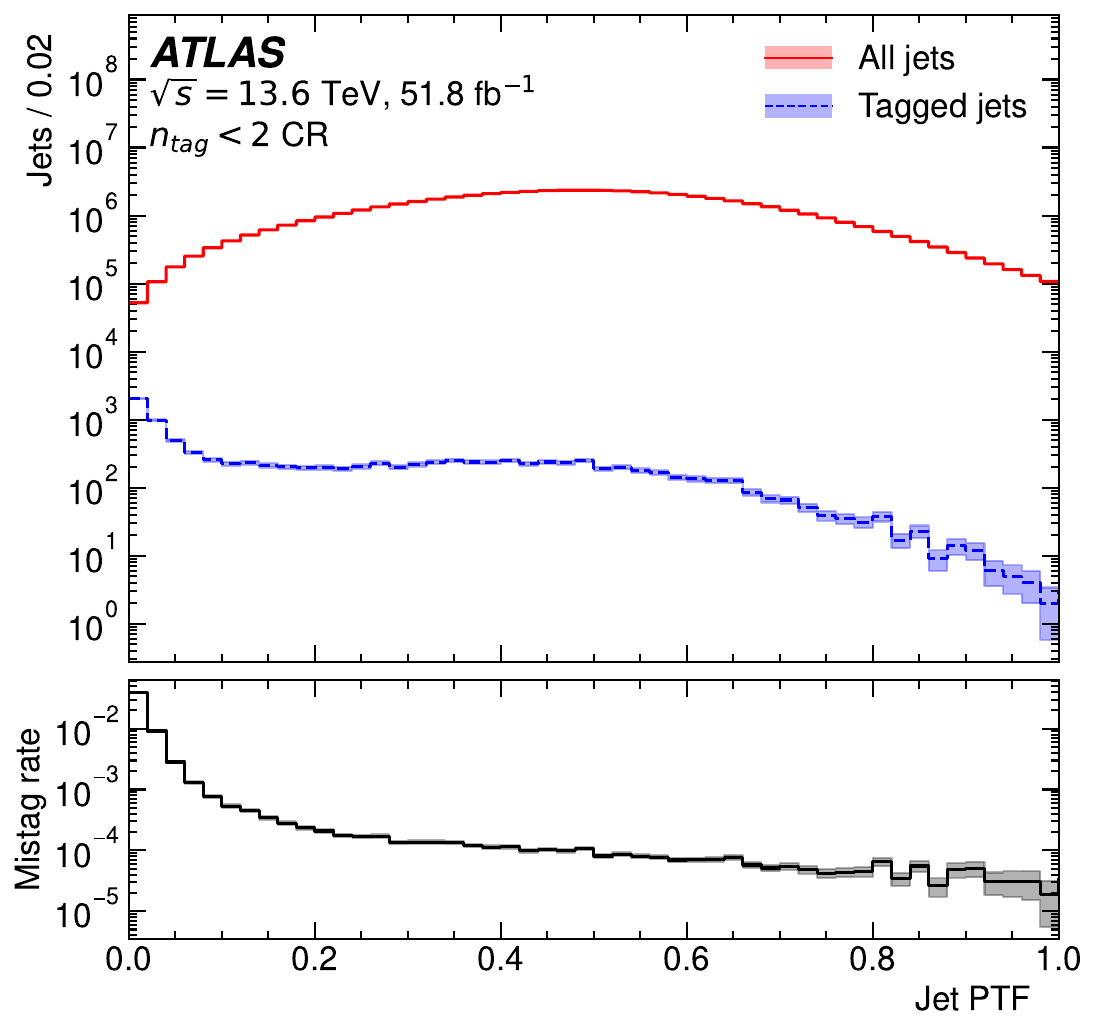}
\caption{}
\label{fig:gnn_mistag_correlation:ptf}
\end{subfigure}  \\
\caption{Distributions of (a) \pt and (b) PTF for all jets (solid lines) and jets with a classification score greater than 0.98 (dashed lines) in the \highmjj selection for data events with $n_{\mathrm{tag}} < 2$. The bottom panel shows the mistag rate as a function of each observable.}
\label{fig:gnn_mistag_correlation}
\end{figure}

To capture these correlations and address the dominant sources of jet--jet correlations, the mistag rates used for both the \lowmjj and \highmjj regions are parameterised based on the observables with the strongest correlations: jet \pt and jet PTF. Figure~\ref{fig:gnn_mistag_correlation} shows the mistag rate as a function of jet \pt and jet PTF, both of which exhibit clear correlations with the mistag rate. The mistag rate is therefore calculated
in 2-dimensional bins of jet \pt and PTF.

Two sources of uncertainty on the background prediction are considered. The first is the statistical uncertainty arising from the finite number of events in the control region used to derive the mistag rates. This is estimated by generating an ensemble of background predictions using a set of $100$ statistically varied per-jet mistag rates, where each bin is randomly shifted according to a Gaussian distribution with a width set by its statistical uncertainty.
The standard deviation of the resulting 100 background estimates defines the statistical uncertainty on the nominal prediction. The statistical uncertainty
computed from this toy method is $6.2\%$ and $2.6\%$ for the \highmjj and \lowmjj
regions, respectively.

The second source of uncertainty is a systematic effect related to the choice of mistag
rate parameterisation. The nominal mistag rate is parameterised using jet \pt and PTF, as these observables are strongly correlated with the transformer jet classification score and account for
potential correlations between jets in an event. However, alternative jet-level observables
exhibiting similar correlations could also be used. To assess the impact of this choice, three
additional parameterisations are considered: the number of $b$-tagged subjets, the number of tracks associated to the jet, and the number of reconstructed secondary vertices associated with the jet using the independent vertex reconstruction algorithm described in Section~\ref{sec:objects}.

To ensure that the dominant \pt and PTF correlations remain accounted for, all alternative mistag
rates include \pt as a parameter and are separated into low-PTF ($<0.04$) and high-PTF ($>0.04$)
categories, resulting in 16 possible parameterisation permutations. The impact of these variations is
evaluated by computing background estimates for each SR using the different
parameterisations. The largest observed variation in the predicted background is taken as the
systematic uncertainty due to the choice of mistag rate parameterisation. This approach quantifies
the effect of potential residual correlations not captured by the nominal parameterisation in \pt and PTF. The resulting systematic uncertainty is 61\% in the \highmjj region and 23\% in the \lowmjj region.

To validate the background estimation strategy, a \textit{validation region tag} is used that defines a ``tag'' as a jet
having a jet classification score between 0.9 and 0.98. This provides a tag definition which captures
jets that are adjacent to the signal tag definition, while removing signal jets, thereby
offering a region to further study the ability of the mistag rate method
to predict yields in a $n_{\mathrm{tag}} \geq 2$ selection. The level of signal
contamination
in the $n_{\mathrm{tag}} \geq 2$ selection using this VR tag definition was checked in simulation and
found to
be below 10\%.

The background estimation strategy described above is applied using the VR tag
definition. Mistag rates are computed following the same procedure as in the nominal
SR, and systematic uncertainties are derived using the same
methodology. The statistical uncertainties are found to be negligible compared to the systematic uncertainties and are not considered. The predicted and observed yields are summarised in
Table~\ref{tab:gnn_VR_results}. The observed yields in the $n_{\mathrm{VR\,tag}} \geq 2$ region are found to be in
agreement with the prediction within uncertainties, demonstrating the robustness of the
background estimation in a signal-adjacent region and indicating that no additional
nonclosure systematic uncertainties are required.
The background estimation strategy was further validated by deriving predictions
from simulated SM multijet events and confirming their agreement with the
observed event yields within uncertainties.
\begin{table}[h]
\centering
\caption{Predicted and observed yields using the validation tag definition
in the \highmjj and \lowmjj selection. The predictions are obtained by summing the event tagging probabilities from Equation~\ref{eq:prob2DV} and \ref{eq:prob1DV}.}
\label{tab:gnn_VR_results}
\begin{tabular}{l|ccc}
\toprule
&  \Lowmjj VR tag & \Highmjj VR tag   \\ \midrule
Pred.  & $174 \pm 42~(\mathrm{syst.})$   & $29 \pm 16~(\mathrm{syst.})$  \\
Obs    & $185$           & $31$          \\
\bottomrule
\end{tabular}
\end{table}

%
%
%


%
\section{Signal systematic uncertainties}
\label{sec:systematics}

Instrumental and theoretical uncertainties are assigned to the modelling of the simulated signal samples.
The experimental uncertainties include contributions from jets, tracks, and luminosity.
The uncertainties associated with large-$R$ jets arise from the jet energy scale (JES)
and jet energy resolution (JER)~\cite{JETM-2018-05}. The constituent small-$R$ jets are
varied within their respective uncertainties, and these variations are
propagated to the large-$R$ jets to quantify their impact on the final signal
yields. The resulting uncertainty on the signal expectations depend mostly on
the mass of the mediator and range between 1\% and 10\% in the
cut-based analysis and between 1\% and 50\% in the ML-based analysis. The larger
uncertainties in the ML-based analysis are driven by samples with a mediator
mass of 600~GeV, where the 300~GeV jet $p_{\mathrm{T}}$ requirement lies near
the kinematic threshold, making the acceptance more sensitive to downward
fluctuations in the jet energy scale.

Uncertainties in track reconstruction are considered for both the primary and large-impact parameter track reconstruction passes. The uncertainty on the efficiency of
large-impact parameter track reconstruction is derived from comparisons between data
and MC simulation in $K^0_S$ events~\cite{IDTR-2021-03}.
For primary tracks, uncertainties are estimated by evaluating the reconstruction
efficiency in alternative simulated samples where the amount of passive material in
the detector is varied~\cite{PERF-2015-07}.
These uncertainties affect track-based observables in the analysis, including the PTF, ECF2, vertex reconstruction, and the transformer jet classification score. To assess their impact, each observable is recomputed using a varied track collection, and the signal yields are rederived. In particular, secondary vertex reconstruction is repeated on the modified track collection to propagate tracking uncertainties to vertex-based observables. The total impact of tracking uncertainties on the final signal yield is estimated to be below 10\%.

To ensure accurate modelling of pile-up conditions, simulated events are reweighted so that
the average number of interactions per bunch crossing matches that observed in data.
Since this number depends on the inelastic $pp$ cross-section, differences between the
predicted and measured values~\cite{STDM-2015-05} are propagated as a systematic
variation of the reweighted distribution, resulting in an uncertainty of the order of 1\% on the signal yield, except in samples with high dark pion lifetime where this uncertainty is
as large as 5\%.

In the cut-based \lowmjj selection, an uncertainty on the trigger scale factor is evaluated by varying the functional form and the parameters of the function modelling the efficiency of the trigger. The envelope of these variations around the nominal scale factor is chosen as the uncertainty. Its impact on the final signal yield is found to be below  10\%.

Finally, the expected number of signal events is subject to an uncertainty in the total
integrated luminosity of the dataset. The combined integrated luminosity for the 2022–2023 dataset has an uncertainty of 2\%, determined using the LUCID-2
detector~\cite{LUCID2}, which provides the primary luminosity measurement.

The MC generator uncertainties applied to the signal include uncertainties in
the PDFs~\cite{Butterworth:2015oua}, renormalisation and factorisation scales,
and the strong coupling constant. Additionally, uncertainties related to the
parton shower tuning parameters in initial- and final-state radiation—such as
variations of the renormalisation scale for QCD emission and the inclusion of
non-singular terms—are considered~\cite{Mrenna:2016sih}. These uncertainties are
quantified by applying event weights corresponding to each variation, allowing
their impact to be propagated through the analysis selections. The combined
impact of these theoretical uncertainties on the signal acceptance is estimated
to be between 2\% and 10\%.

The leading source of uncertainty varies across the signal parameter space, with
jet-related uncertainties dominating in $s$-channel scenarios with $m_{Z'} = 600$~GeV,
and theoretical modelling uncertainties dominating at higher mediator masses.


%
\section{Results}
\label{sec:results}

The predicted event yields (as obtained from the background-only ABCD fit or the mistag rate methods discussed in sections~\ref{sec:bkg_cutbased} and \ref{sec:bkg_gnn}) and the  corresponding observed yields for both analysis strategies are summarised in Table~\ref{tab:results}. No significant excess above the expected background is
observed.

\begin{table}[t!]
\centering
\caption{The predicted number of background events and the observed event yields, including statistical and systematic uncertainties in the background estimation. The
cut-based predictions are computed from the background-only fit.}
\renewcommand{\arraystretch}{1.5} %
\begin{tabular}{ccccc|c}
\toprule
\multicolumn{1}{c|}{Strategy} & \multicolumn{1}{c|}{Region} & \multicolumn{3}{c|}{Prediction ($\pm$ stat $\pm$ syst)} & Observed yield        \\ \hline
\multicolumn{1}{c|}{\multirow{2}{*}{Cut-based}} & \multicolumn{1}{c|}{High-$m_{jj}$} & $7.5$                  & $\pm 1.1$                  & $\pm 1.1$                 & \multicolumn{1}{c}{8} \\ \cline{2-6}
\multicolumn{1}{c|}{}                                         & \multicolumn{1}{c|}{Low-$m_{jj}$}  & $16.6$            & $\pm 4.8$                  & $\pm 4.8$                 & \multicolumn{1}{c}{10} \\ \hline \hline
\multicolumn{1}{c|}{\multirow{2}{*}{ML-based}}       & \multicolumn{1}{c|}{High-$m_{jj}$} & $4.5$             & $\pm 0.3$                  &  $\pm 2.8$                & \multicolumn{1}{c}{3} \\ \cline{2-6}
\multicolumn{1}{c|}{}                                         & \multicolumn{1}{c|}{Low-$m_{jj}$}  & $31.8$            & $\pm 0.8$                  & $\pm 7.5$                 & \multicolumn{1}{c}{24} \\ \bottomrule
\end{tabular}
\label{tab:results}
\end{table}

Two separate statistical treatments are performed, one for the cut-based
analysis and another for the ML-based analysis. Within each approach, a simultaneous
likelihood fit is performed, combining the low-\(m_{jj}\) and high-\(m_{jj}\) signal
regions. Systematic uncertainties on the signal and background predictions are
incorporated as nuisance parameters in both likelihood functions.

In the cut-based analysis, the background estimation follows the ABCD method separately in the \lowmjj and \highmjj regions. An overall profile likelihood
function is constructed from the product of the Poisson probabilities of observing
the number of events $N_{\text{obs}, X}$, given an expectation $N_{\text{exp}, X}$, in each region $X \in \{A, B, C, D\}$.
The expected  yield in each region is expressed as $N_{\text{pred},X} = \mu N_{\text{sig},X}+N_{\text{bkg}, X}$  where $N_{\text{sig}, X}$, is the number of expected signal events entering region $X$, $\mu$ is the signal strength  and $N_{\text{bkg}, X}$ the expected background yield in this region.   The background yields in each ABCD plane are expressed using three free parameters,  $m$, $\tau_B$ and $\tau_C$, as :
\begin{align*}
N_{\text{bkg},A} &= m \\
N_{\text{bkg},B} &= m \tau_B \\
N_{\text{bkg},C} &= m \tau_C \\
N_{\text{bkg},D} &= m \tau_B \tau_C
\end{align*}
therefore automatically imposing the ABCD relation of Equation~\ref{eq:abcd_pred}.
The combined fit to both \lowmjj and \highmjj regions thus involves seven free parameters, including $\mu$ which is common to both regions.

For the cut-based search, the combined likelihood function is
\begin{equation}
\mathcal{L}(\mu, \theta) = \prod_{r \in \{\lowmjj, \highmjj\}} \prod_{X \in \{A, B, C, D\}} P(N_{\text{obs}, X}^{(r)} \mid N_{\text{exp}, X}^{(r)}(\mu, \theta)) \times \prod_{i} G(\theta_i \mid 0, 1),
\end{equation}
where $P(N_{\text{obs}, X}^{(r)} \mid N_{\text{exp}, X}^{(r)}(\mu, \theta))$
represents the Poisson probability of observing $N_{\text{obs}, X}^{(r)}$ events
given the expected yield $N_{\text{exp}, X}^{(r)}$, which includes both background
and signal contributions, and $G(\theta_i \mid 0,1)$ denotes Gaussian constraints on
nuisance parameters $\theta_i$.

Under the background-only hypothesis, the expected background yields are
initially set to their observed values, defining the \textit{a priori}
background estimate. When the observed yields in the signal regions are
included, the fit allows the expected background to adjust, resulting in the
\textit{a posteriori} background estimate. The \textit{a posteriori} estimates differ by less than 1\% from the \textit{a priori} values and lie well within the statistical uncertainties of the fit.
In the presence of a signal, the expected background contribution is modified dynamically, enabling
an excess to be interpreted as a nonzero signal strength.

In the ML-based search, the likelihood function follows the same structure as in
the cut-based analysis, but includes only two Poisson terms: one for the
observed event yield in each of the \lowmjj and \highmjj signal regions.

Upper limits at the 95\% confidence level (CL) are set on the production cross-section
times branching fraction for each signal hypothesis using the
CL\textsubscript{s} method~\cite{Read:2002hq}. The profile likelihood ratio is
used as the test statistic, with its distribution approximated using asymptotic
formulae~\cite{Cowan:2010js}. The validity of this approximation is confirmed by
comparison with a full frequentist pseudo-experiment-based method across a range of
signal samples, yielding consistent results. The likelihood function is implemented
using the \texttt{pyhf} framework~\cite{pyhf, pyhf_joss}.

\begin{figure}[h]
\centering
\begin{subfigure}{0.475\textwidth}
\centering
\includegraphics[width=\textwidth]{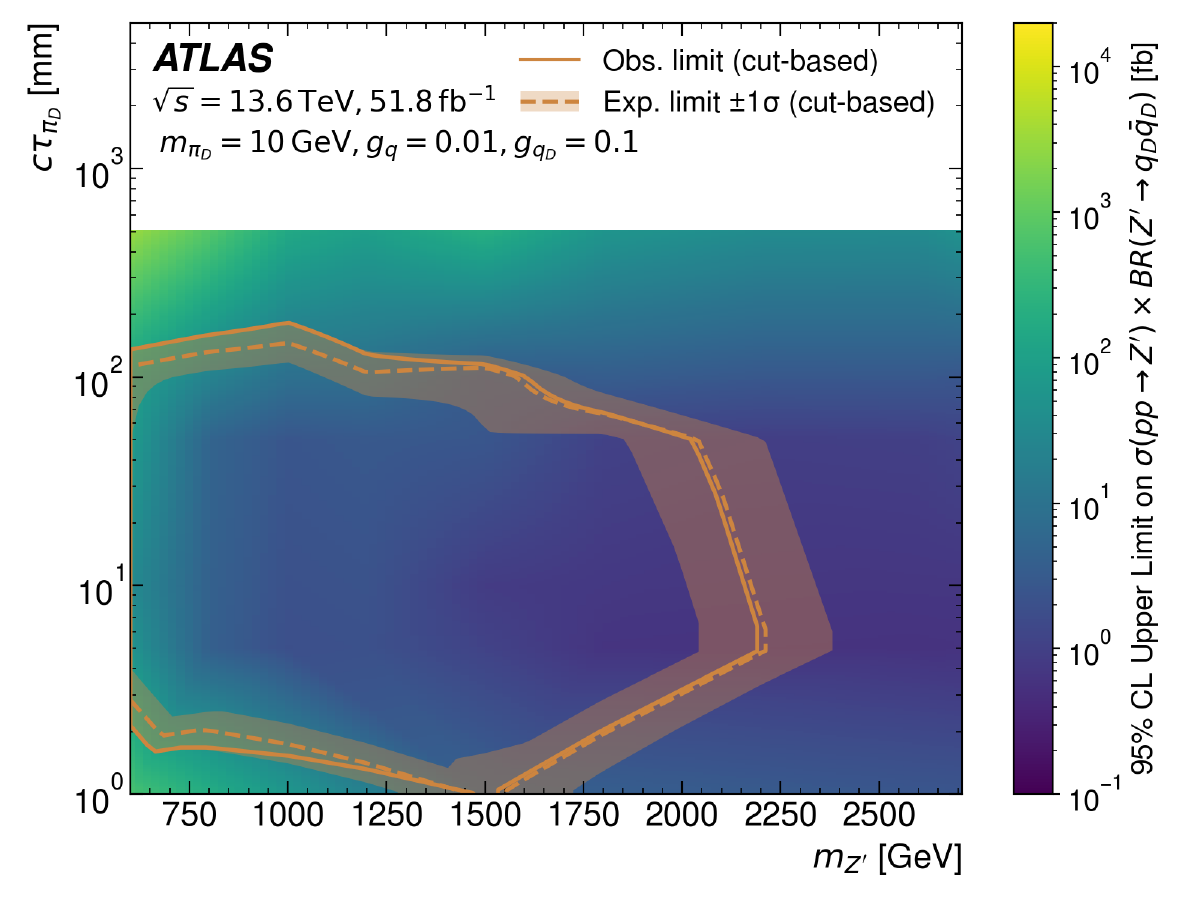}
\label{fig:limits_gnn_ML_ABCD:ABCD}
\caption{}
\end{subfigure}
\begin{subfigure}{0.475\textwidth}
\centering
\includegraphics[width=\textwidth]{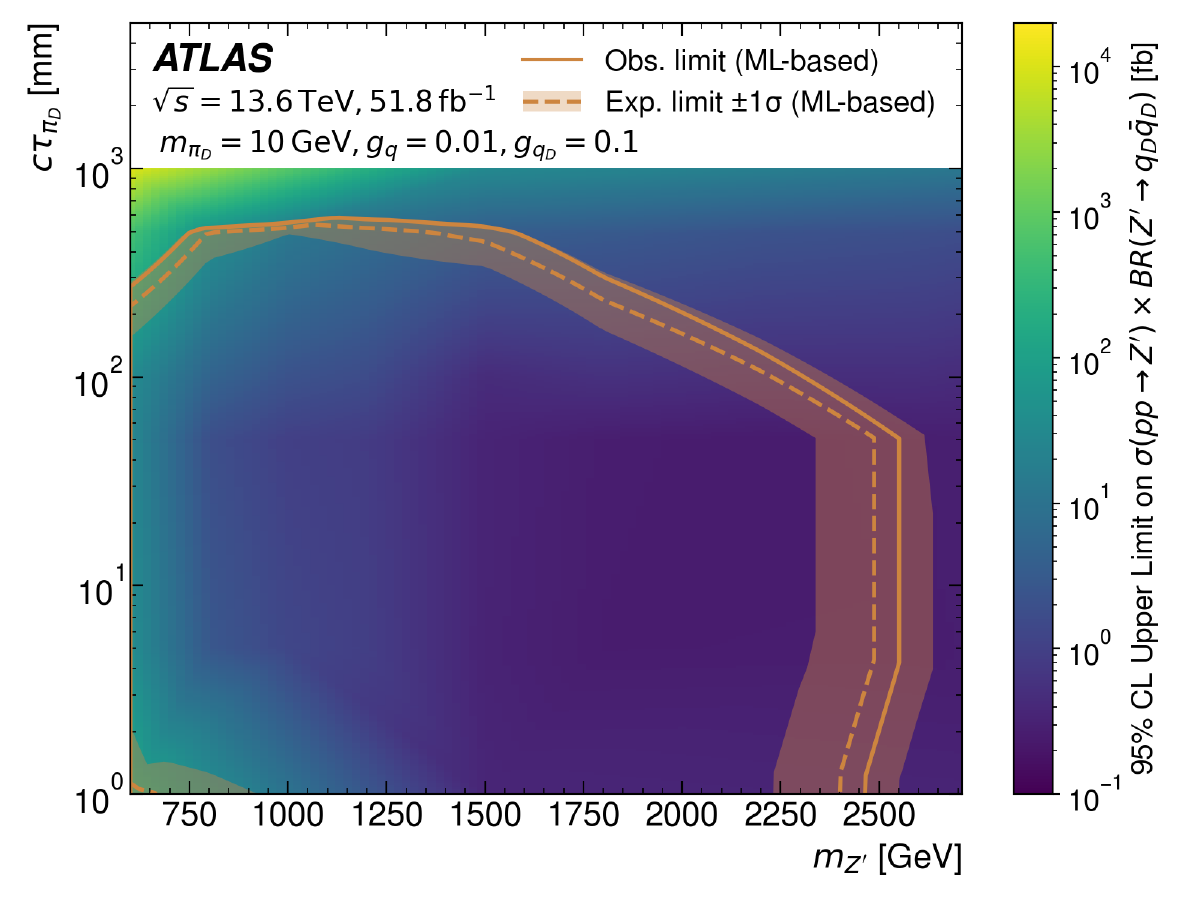}
\label{fig:limits_gnn_ML_ABCD:ML}
\caption{}
\end{subfigure}
\caption{The 95\% CL exclusion limits on $\sigma(pp \rightarrow Z')
\times \mathrm{Br}(Z' \rightarrow q_D \bar{q}_D)$ as a function of $m_{Z'}$ and
$c\tau_{\pi_D}$ for (a) the cut-based analysis and (b) the ML-based analysis. The excluded region computed from the observed limits for $g_q = 0.01$ and $g_{q_D} = 0.1$
is shown as a solid line.}
\label{fig:limits_gnn_ML_ABCD}
\end{figure}

The expected and observed limits on $\sigma(pp \rightarrow Z') \times \mathrm{Br}(Z' \rightarrow q_D \bar{q}_D)$ are shown
in Figure~\ref{fig:limits_gnn_ML_ABCD} as functions of $m_{Z'}$ and
$c\tau_{\pi_D}$ for both the ML-based and cut-based strategies. The ML-based
strategy sets the most stringent limits across the benchmark models studied.
Assuming coupling values of $g_q = 0.01$ and $g_{q_D} = 0.1$, the ML-based
analysis excludes $Z'$ masses up to 2550~\GeV for $c\tau_{\pi_D} =
10~\mathrm{mm}$. Under the same assumptions, the cut-based analysis excludes
$Z'$ masses up to 2150~\GeV. For $m_{Z'} = 1000~\mathrm{\GeV}$, the ML-based
strategy excludes dark pion lifetimes in the range 1--500~mm, while the
cut-based strategy excludes lifetimes between 1.5~mm and 200~mm.
The impact of systematic uncertainties on the observed limits is typically at
the percent level, with the exception of signal samples with $m_{Z'} = 600$~GeV
where larger jet-related uncertainties in the ML-based analysis lead to a
degradation in sensitivity of approximately 20\%.

For $m_{Z'} = 1500~\GeV$ and $c\tau_{\pi_D} = 50$~mm, the ML-based analysis
excludes values of $g_q > 0.003$ assuming $g_{q_D} > 0.03$. This coupling
strength is more than 20 times smaller than the limits set by dijet resonance
searches\cite{EXOT-2019-03,EXOT-2018-62}, demonstrating the power of the
targeted emerging jet analysis strategy to probe previously inaccessible regions
of parameter space.

\begin{figure}[h!]
\centering
\begin{subfigure}[b]{0.475\textwidth}
\centering
\includegraphics[width=\textwidth]{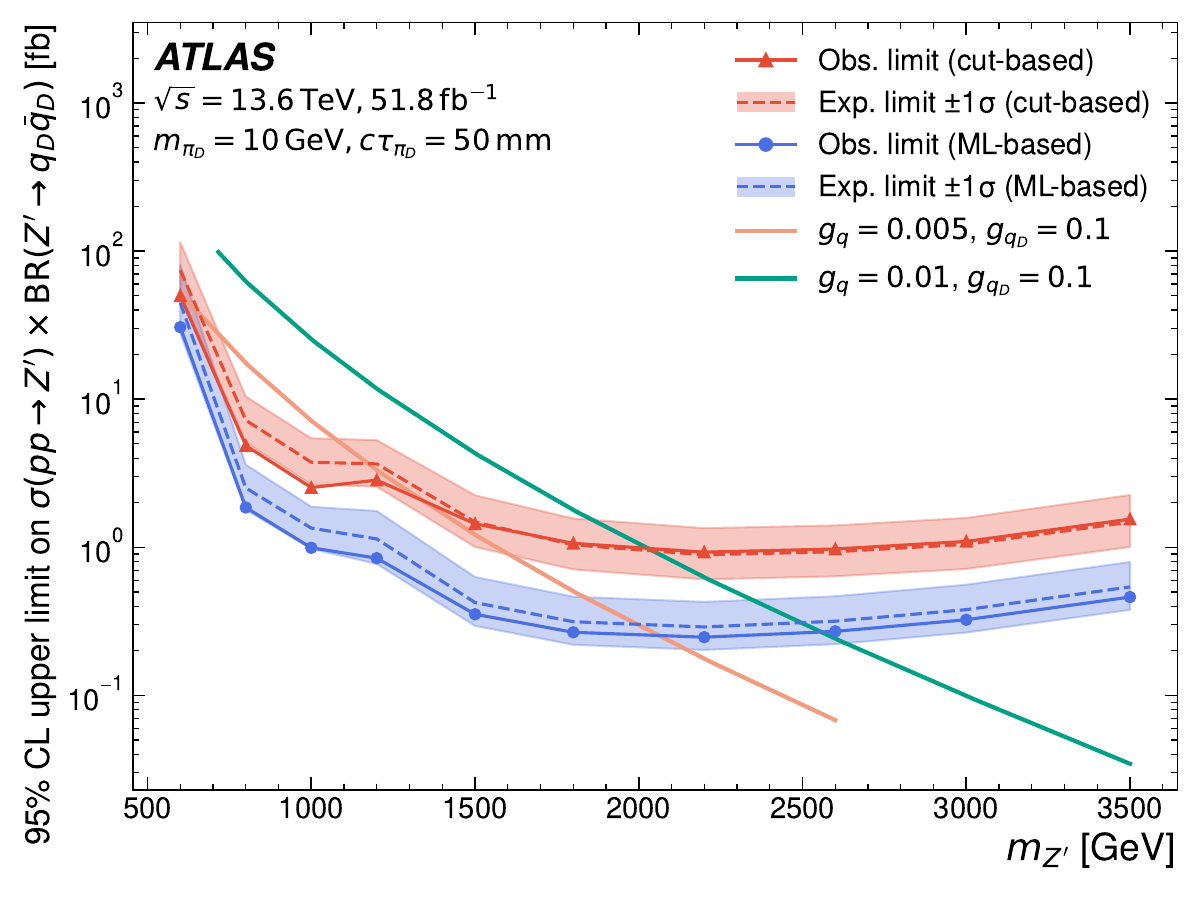}
\caption{}
\label{fig:limits_gnn_combined:sigma_vs_mZ_bands_ctau50}
\end{subfigure}
\begin{subfigure}[b]{0.475\textwidth}
\centering
\includegraphics[width=\textwidth]{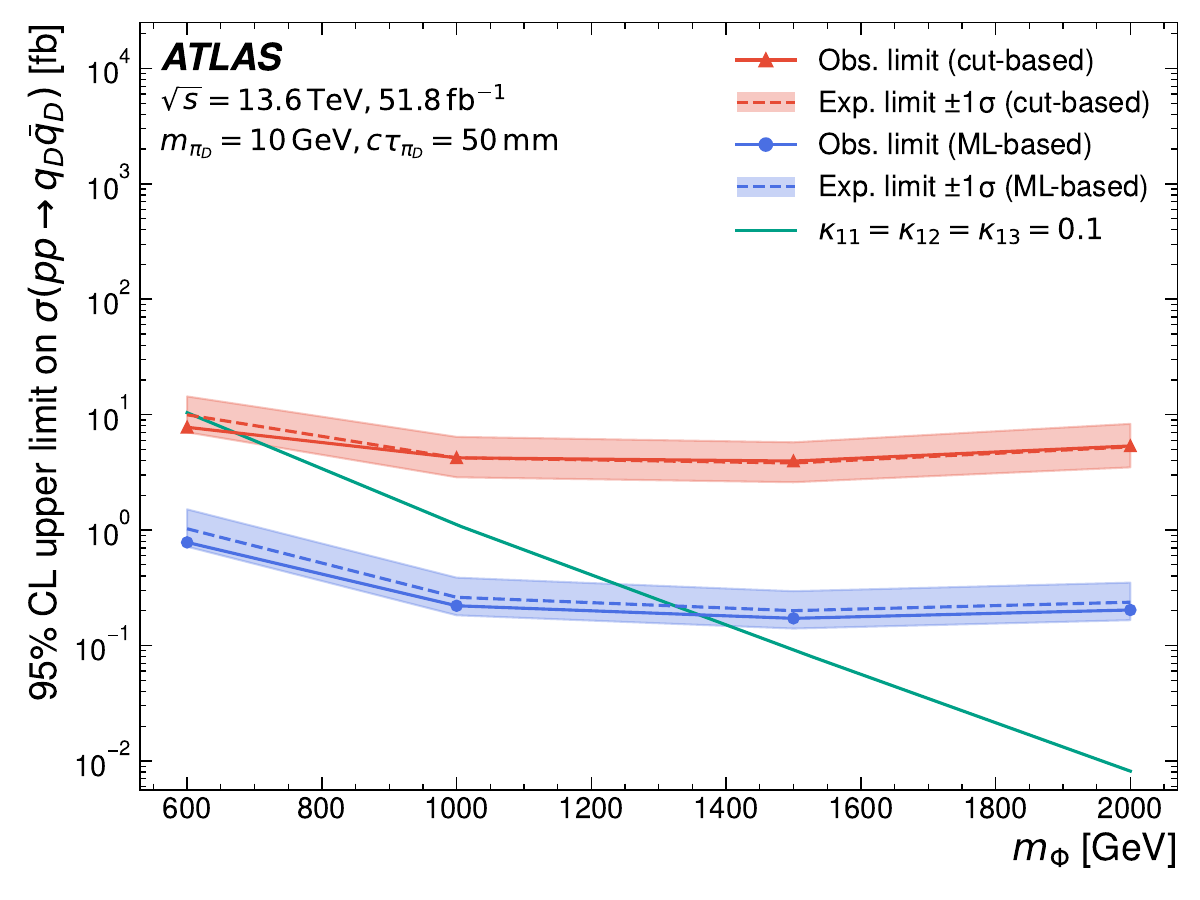}
\caption{}
\label{fig:limits_gnn_combined:sigma_vs_mXd_bands_ctau50}
\end{subfigure}
\begin{subfigure}[b]{0.475\textwidth}
\centering
\includegraphics[width=\textwidth]{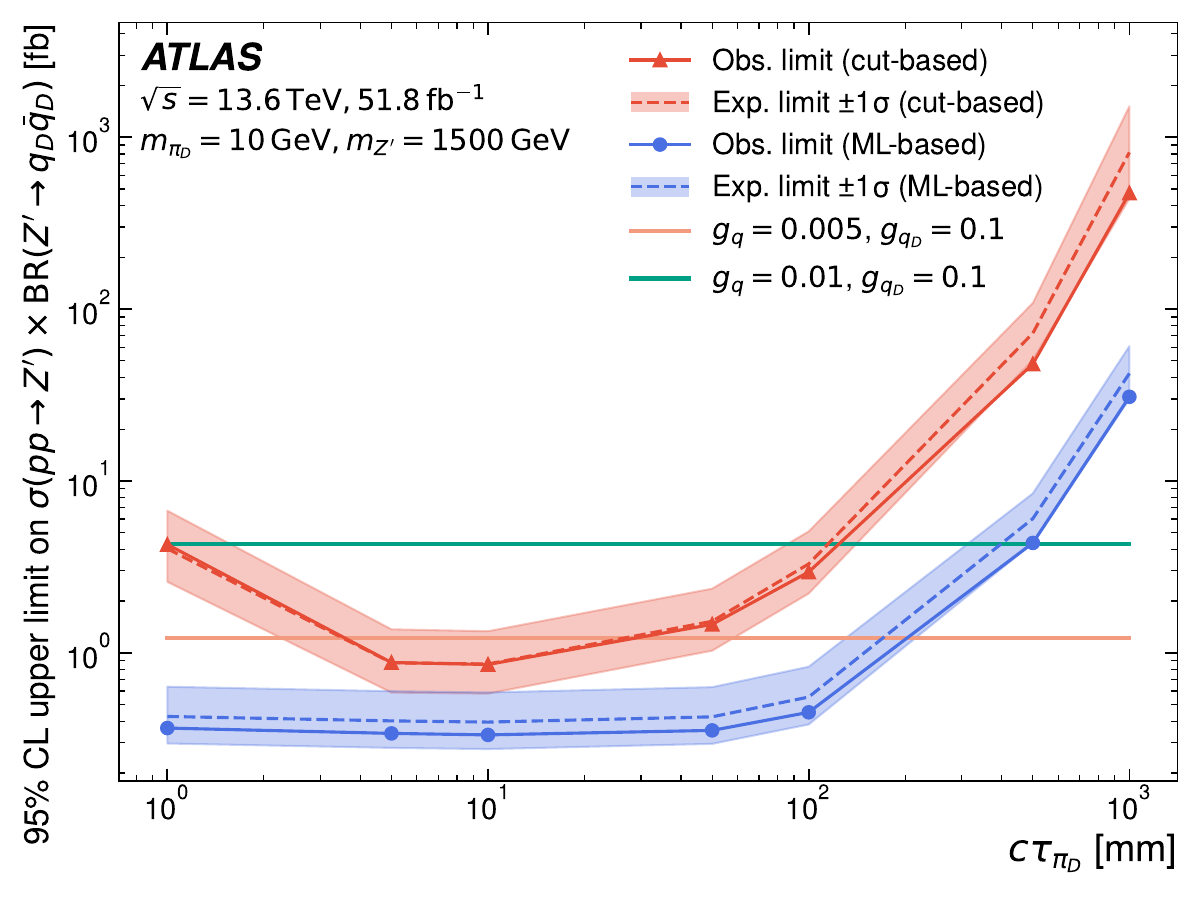}
\caption{}
\label{fig:limits_gnn_combined:sigma_vs_ctau_bands_1500}
\end{subfigure}
\begin{subfigure}[b]{0.475\textwidth}
\centering
\includegraphics[width=\textwidth]{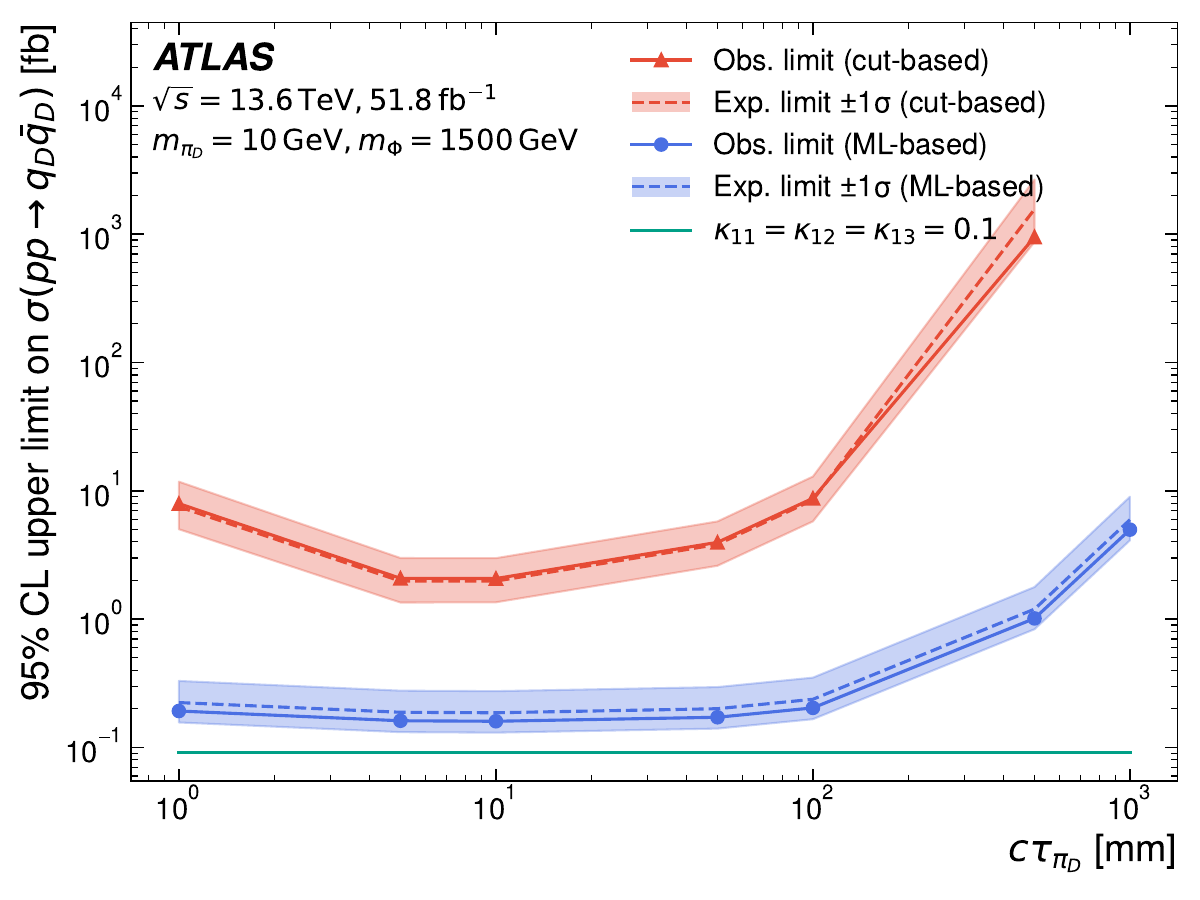}
\caption{}
\label{fig:limits_gnn_combined:sigma_vs_ctau_bands_1500_tchan}
\end{subfigure}
\caption{The 95\% CL exclusion limits on (a, c) $\sigma(pp \rightarrow Z')
\times \mathrm{Br}(Z' \rightarrow q_D \bar{q}_D)$ and (b, d) $\sigma(pp \rightarrow q_D \bar{q}_D)$
for  $m_{\pi_D}=10$~\GeV for the cut-based (triangles) and ML-based (circles) strategies. The top row displays the limits as a function of $m_{Z',\Phi}$ for fixed $c\tau_{\pi_D} = 50$~mm, while the bottom row displays the limits as a function of $c\tau_{\pi_D}$ for $m_{Z',\Phi} = 1500$~\GeV. Theoretical predictions
are shown as solid lines.}
\label{fig:limits_gnn_combined}
\end{figure}

Figure~\ref{fig:limits_gnn_combined} shows the 95\% CL upper limits as a
function of either $m_{Z',\Phi}$ or $c\tau_{\pi_D}$ assuming $m_{\pi_D} = 10$~\GeV.
In the $s$-channel scenario, the
limits are strongest for $m_{Z'} > 1200$~\GeV, where the \highmjj region
becomes the dominant contribution to the signal acceptance. In the $t$-channel scenario, there is a reduced dependence on the mediator mass due to the weaker
correlation between $m_{\Phi}$ and $m_{jj}$. In the ML-based
strategy, the limits for both benchmark scenarios remain largely constant for $1~\mathrm{mm} < c\tau_{\pi_D}
< 100~\mathrm{mm}$, but weaken for $c\tau_{\pi_D} > 100~\mathrm{mm}$ due to
reduced track and jet reconstruction efficiency at larger radial decay positions. In
contrast, the limits obtained from the cut-based analysis weaken at lower dark
pion lifetimes due to the explicit requirements placed on PTF and
$N_{\mathrm{vtx}}$. For the $t$-channel models, the limits set by the ML-based
analysis are more than an order of magnitude stronger than those set by the
cut-based analysis for $c\tau_{\pi_D} = 50$~mm, highlighting the ability of the
transformer-based tagger to generalise to physics processes that were not
considered in the network training.

\begin{figure}[h!]
\centering
\begin{subfigure}[b]{0.475\textwidth}
\centering
\includegraphics[width=\textwidth]{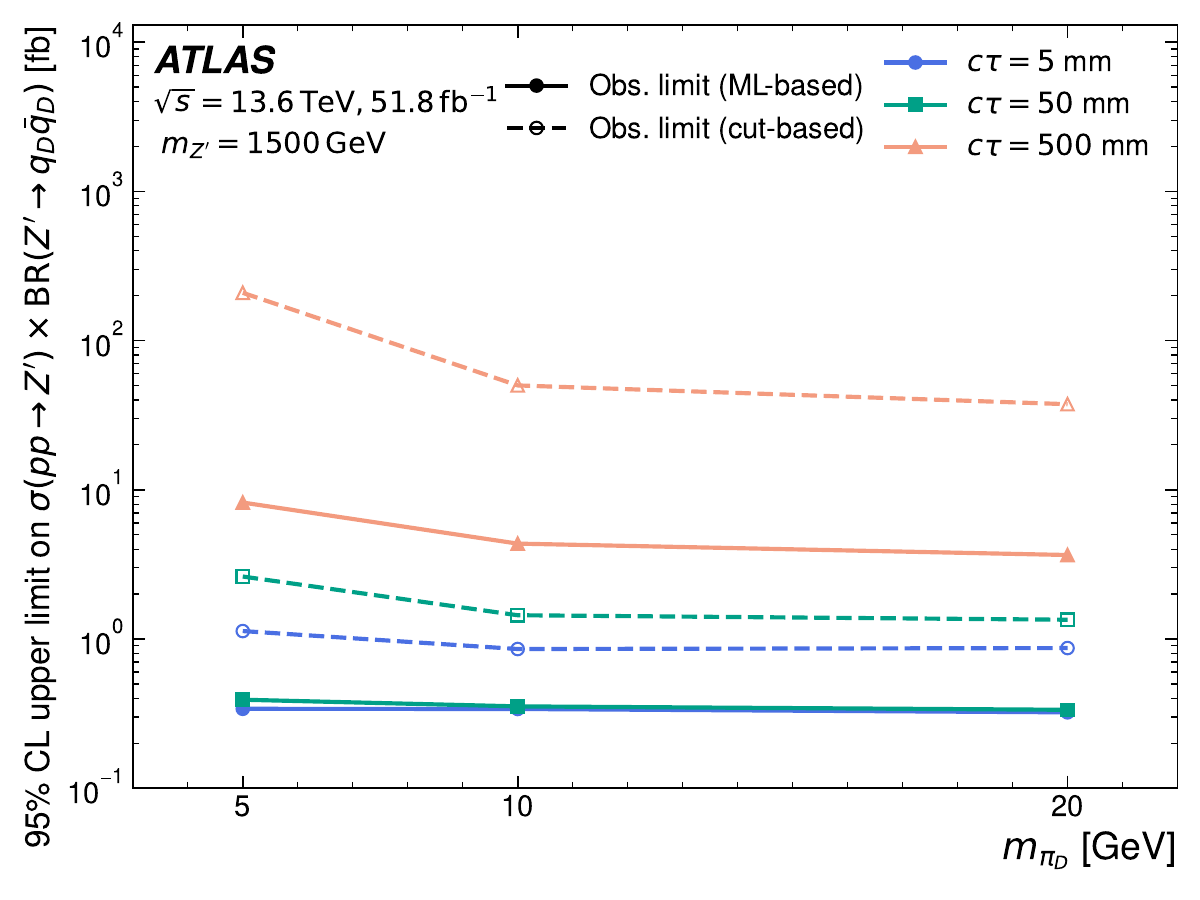}
\caption{}
\label{fig:limits_gnn_mPi:sigma_vs_piDmass_1500}
\end{subfigure}
\begin{subfigure}[b]{0.475\textwidth}
\centering
\includegraphics[width=\textwidth]{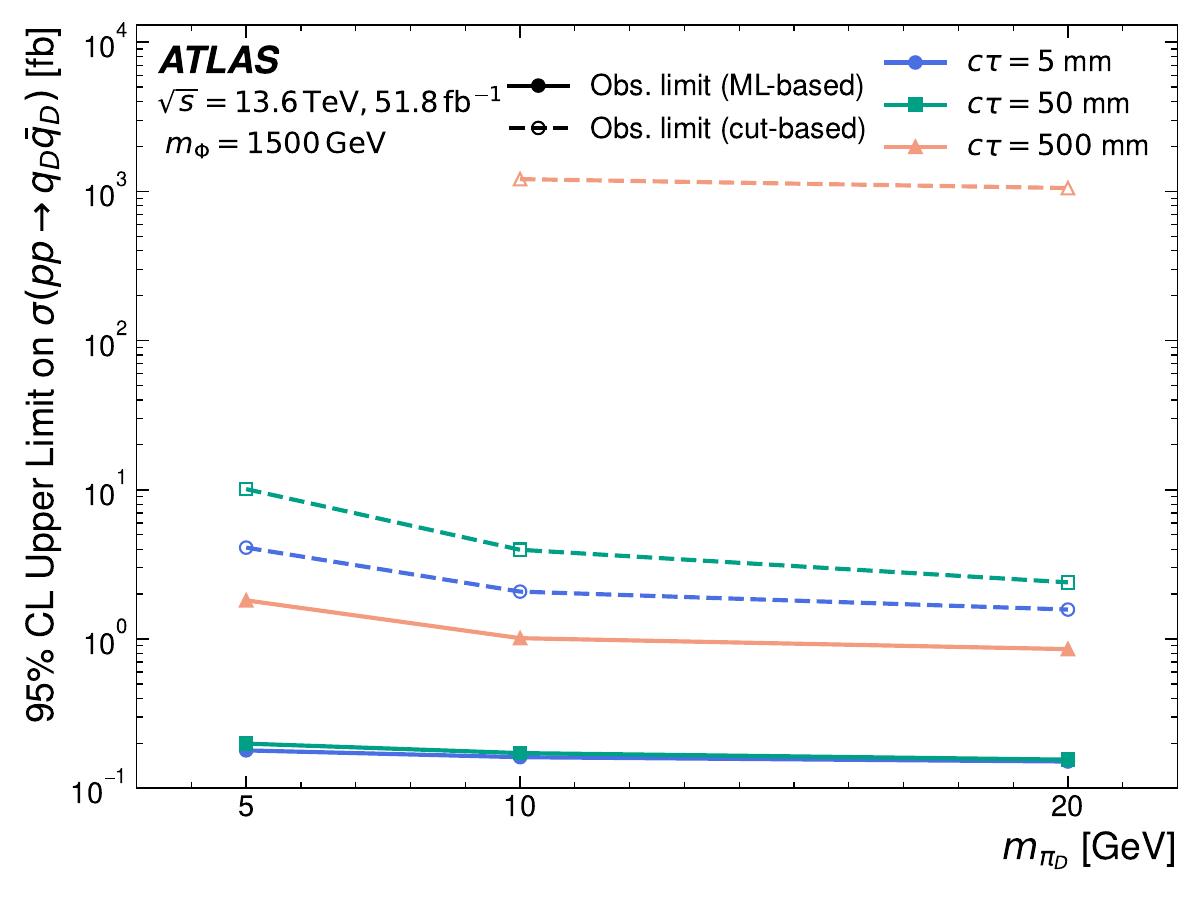}
\caption{}
\label{fig:limits_gnn_mPi:sigma_vs_piDmass_1500_tchan}
\end{subfigure}
\caption{The 95\% CL exclusion limits on (a) $\sigma(pp \rightarrow Z')
\times \mathrm{Br}(Z' \rightarrow q_D \bar{q}_D)$ and (b) $\sigma(pp \rightarrow
q_D \bar{q}_D)$ for $m_{Z',\Phi} = 1500$~\GeV for the cut-based (hollow markers and dashed lines) and ML-based (filled markers and solid lines) strategies. The
limits are shown for $c\tau_{\pi_D} = 5$~mm (circles), $50$~mm (squares) and $500$~mm (triangles).
The point corresponding to the cut-based analysis for $m_{\pi_D} = 5$~GeV and
$c\tau_{\pi_D} = 500$~mm in (b) is not shown due to insufficient signal efficiency
to derive a valid exclusion limit.}
\label{fig:limits_gnn_mPi}
\end{figure}

Figure~\ref{fig:limits_gnn_mPi} shows the observed upper limits as a function of
$m_{\pi_D}$. Both the cut-based and ML-based strategies exhibit minimal
dependence on the dark pion mass, with the exception of
dark pion proper decay lengths of 500~mm. As described in Section~\ref{sec:analysis},
the increased boost of lighter dark pions causes their decay vertices to occur
further from the interaction point, often outside the acceptance of the inner
detector, leading to a degradation in sensitivity.
While the impact of the specific choices of $N_c$, $N_f$, and the mass hierarchy
$m_{\rho_D} = 2\Lambda_D = 2 m_{q_D} = 4 m_{\pi_D}$ on the analysis sensitivity has
not been explicitly studied, these parameters are primarily expected to affect
the dark pion multiplicity~\cite{Schwaller:2015gea}. Although this can influence the
overall signal acceptance, it is not expected to significantly modify the key
features of the final-state signature targeted by the analysis. The results
presented here are therefore expected to remain sensitive across a broader range
of dark QCD parameter choices.

%
%
%


%
\section{Conclusion}
\label{sec:conclusion}

This paper presents a search for emerging jets using $51.8~\mathrm{fb}^{-1}$ of
$\sqrt{s} = 13.6$~\TeV $pp$ collision data collected by the ATLAS experiment at
the LHC between 2022 and 2023. Two complementary analysis strategies are
employed: one based on selections of high-level jet observables and the other
using a custom transformer-based emerging jet tagging algorithm. The
analysis is performed separately in the \lowmjj and \highmjj regions, each with
distinct trigger strategies. No significant excess is observed above the
expected background, which is estimated using a fully data-driven approach.
Limits are placed at the 95\% confidence level on two different mediator
scenarios: a vector mediator $Z'$ and a scalar bi-fundamental mediator $\Phi$.
$Z'$ masses of up to 2.5 \TeV are excluded assuming quark and dark quark coupling
values of $g_q = 0.01$ and $g_{q_D} = 0.1$, respectively, and $\Phi$ masses up to 1350~\GeV are excluding for a quark-dark quark coupling value of 0.1.
This search is part
of a broader search programme for dark QCD signatures within the ATLAS experiment
and offers unique sensitivity to dark mesons with laboratory decay lengths on
the order of 100~mm. Notably, it represents the first search for emerging jet
production via a resonant $s$-channel mediator and the first application of a
transformer-based algorithm for emerging jet tagging.
This analysis establishes new constraints on emerging jet production and
represents a significant step forward in the ATLAS dark sector search programme.


%
%
%
%
%
\section*{Acknowledgements}
%

%

%
%

%
%

We thank CERN for the very successful operation of the LHC and its injectors, as well as the support staff at
CERN and at our institutions worldwide without whom ATLAS could not be operated efficiently.

The crucial computing support from all WLCG partners is acknowledged gratefully, in particular from CERN, the ATLAS Tier-1 facilities at TRIUMF/SFU (Canada), NDGF (Denmark, Norway, Sweden), CC-IN2P3 (France), KIT/GridKA (Germany), INFN-CNAF (Italy), NL-T1 (Netherlands), PIC (Spain), RAL (UK) and BNL (USA), the Tier-2 facilities worldwide and large non-WLCG resource providers. Major contributors of computing resources are listed in Ref.~\cite{ATL-SOFT-PUB-2025-001}.

We gratefully acknowledge the support of ANPCyT, Argentina; YerPhI, Armenia; ARC, Australia; BMWFW and FWF, Austria; ANAS, Azerbaijan; CNPq and FAPESP, Brazil; NSERC, NRC and CFI, Canada; CERN; ANID, Chile; CAS, MOST and NSFC, China; Minciencias, Colombia; MEYS CR, Czech Republic; DNRF and DNSRC, Denmark; IN2P3-CNRS and CEA-DRF/IRFU, France; SRNSFG, Georgia; BMFTR, HGF and MPG, Germany; GSRI, Greece; RGC and Hong Kong SAR, China; ICHEP and Academy of Sciences and Humanities, Israel; INFN, Italy; MEXT and JSPS, Japan; CNRST, Morocco; NWO, Netherlands; RCN, Norway; MNiSW, Poland; FCT, Portugal; MNE/IFA, Romania; MSTDI, Serbia; MSSR, Slovakia; ARIS and MVZI, Slovenia; DSI/NRF, South Africa; MICIU/AEI, Spain; SRC and Wallenberg Foundation, Sweden; SERI, SNSF and Cantons of Bern and Geneva, Switzerland; NSTC, Taipei; TENMAK, T\"urkiye; STFC/UKRI, United Kingdom; DOE and NSF, United States of America.

Individual groups and members have received support from BCKDF, CANARIE, CRC and DRAC, Canada; CERN-CZ, FORTE and PRIMUS, Czech Republic; COST, ERC, ERDF, Horizon 2020, ICSC-NextGenerationEU and Marie Sk{\l}odowska-Curie Actions, European Union; Investissements d'Avenir Labex, Investissements d'Avenir Idex and ANR, France; DFG and AvH Foundation, Germany; Herakleitos, Thales and Aristeia programmes co-financed by EU-ESF and the Greek NSRF, Greece; BSF-NSF and MINERVA, Israel; NCN and NAWA, Poland; La Caixa Banking Foundation, CERCA Programme Generalitat de Catalunya and PROMETEO and GenT Programmes Generalitat Valenciana, Spain; G\"{o}ran Gustafssons Stiftelse, Sweden; The Royal Society and Leverhulme Trust, United Kingdom.

In addition, individual members wish to acknowledge support from CERN: European Organization for Nuclear Research (CERN DOCT); Chile: Agencia Nacional de Investigaci\'on y Desarrollo (FONDECYT 1230812, FONDECYT 1240864); China: Chinese Ministry of Science and Technology (MOST-2023YFA1605700, MOST-2023YFA1609300), National Natural Science Foundation of China (NSFC - 12175119, NSFC 12275265); Czech Republic: Czech Science Foundation (GACR - 24-11373S), Ministry of Education Youth and Sports (ERC-CZ-LL2327, FORTE CZ.02.01.01/00/22\_008/0004632), PRIMUS Research Programme (PRIMUS/21/SCI/017); EU: H2020 European Research Council (ERC - 101002463); European Union: European Research Council (BARD No. 101116429, ERC - 948254, ERC 101089007), European Regional Development Fund (SMASH COFUND 101081355, SLO ERDF), Horizon 2020 Framework Programme (MUCCA - CHIST-ERA-19-XAI-00), European Union, Future Artificial Intelligence Research (FAIR-NextGenerationEU PE00000013), Italian Center for High Performance Computing, Big Data and Quantum Computing (ICSC, NextGenerationEU); France: Agence Nationale de la Recherche (ANR-21-CE31-0022, ANR-22-EDIR-0002); Germany: Baden-Württemberg Stiftung (BW Stiftung-Postdoc Eliteprogramme), Deutsche Forschungsgemeinschaft (DFG - 469666862, DFG - CR 312/5-2); China: Research Grants Council (GRF); Italy: Istituto Nazionale di Fisica Nucleare (ICSC, NextGenerationEU), Ministero dell'Università e della Ricerca (NextGenEU 153D23001490006 M4C2.1.1, NextGenEU I53D23000820006 M4C2.1.1, NextGenEU I53D23001490006 M4C2.1.1, SOE2024\_0000023); Japan: Japan Society for the Promotion of Science (JSPS KAKENHI JP22H01227, JSPS KAKENHI JP22H04944, JSPS KAKENHI JP22KK0227, JSPS KAKENHI JP24K23939, JSPS KAKENHI JP25H00650, JSPS KAKENHI JP25H01291, JSPS KAKENHI JP25K01023); Norway: Research Council of Norway (RCN-314472); Poland: Ministry of Science and Higher Education (IDUB AGH, POB8, D4 no 9722), Polish National Science Centre (NCN 2021/42/E/ST2/00350, NCN OPUS 2023/51/B/ST2/02507, NCN OPUS nr 2022/47/B/ST2/03059, NCN UMO-2019/34/E/ST2/00393, UMO-2022/47/O/ST2/00148, UMO-2023/49/B/ST2/04085, UMO-2023/51/B/ST2/00920, UMO-2024/53/N/ST2/00869); Portugal: Foundation for Science and Technology (FCT); Spain: Ministry of Science and Innovation (MCIN \& NextGenEU PCI2022-135018-2, MICIN \& FEDER PID2021-125273NB, RYC2019-028510-I, RYC2020-030254-I, RYC2021-031273-I, RYC2022-038164-I); Sweden: Carl Trygger Foundation (Carl Trygger Foundation CTS 22:2312), Swedish Research Council (Swedish Research Council 2023-04654, VR 2021-03651, VR 2022-03845, VR 2022-04683, VR 2023-03403, VR 2024-05451), Knut and Alice Wallenberg Foundation (KAW 2018.0458, KAW 2022.0358, KAW 2023.0366); Switzerland: Swiss National Science Foundation (SNSF - PCEFP2\_194658); United Kingdom: Leverhulme Trust (Leverhulme Trust RPG-2020-004), Royal Society (NIF-R1-231091); United States of America: U.S. Department of Energy (ECA DE-AC02-76SF00515), Neubauer Family Foundation.

%
%


%
\clearpage
\appendix

\printbibliography
\clearpage
 
\begin{flushleft}
\hypersetup{urlcolor=black}
{\Large The ATLAS Collaboration}

\bigskip

\AtlasOrcid[0000-0002-6665-4934]{G.~Aad}$^\textrm{\scriptsize 104}$,
\AtlasOrcid[0000-0001-7616-1554]{E.~Aakvaag}$^\textrm{\scriptsize 17}$,
\AtlasOrcid[0000-0002-5888-2734]{B.~Abbott}$^\textrm{\scriptsize 123}$,
\AtlasOrcid[0000-0002-0287-5869]{S.~Abdelhameed}$^\textrm{\scriptsize 119a}$,
\AtlasOrcid[0000-0002-1002-1652]{K.~Abeling}$^\textrm{\scriptsize 55}$,
\AtlasOrcid[0000-0001-5763-2760]{N.J.~Abicht}$^\textrm{\scriptsize 49}$,
\AtlasOrcid[0000-0002-8496-9294]{S.H.~Abidi}$^\textrm{\scriptsize 30}$,
\AtlasOrcid[0009-0003-6578-220X]{M.~Aboelela}$^\textrm{\scriptsize 45}$,
\AtlasOrcid[0000-0002-9987-2292]{A.~Aboulhorma}$^\textrm{\scriptsize 36e}$,
\AtlasOrcid[0000-0001-5329-6640]{H.~Abramowicz}$^\textrm{\scriptsize 157}$
% ...
\bigskip
\\

% ...
$^{17}$Department for Physics and Technology, University of Bergen, Bergen; Norway.\\
% ...
$^{30}$Physics Department, Brookhaven National Laboratory, Upton NY; United States of America.\\
% ...
$^{36}$$^{(a)}$Facult\'e des Sciences Ain Chock, Universit\'e Hassan II de Casablanca;$^{(b)}$Facult\'{e} des Sciences, Universit\'{e} Ibn-Tofail, K\'{e}nitra;$^{(c)}$Facult\'e des Sciences Semlalia, Universit\'e Cadi Ayyad, LPHEA-Marrakech;$^{(d)}$LPMR, Facult\'e des Sciences, Universit\'e Mohamed Premier, Oujda;$^{(e)}$Facult\'e des sciences, Universit\'e Mohammed V, Rabat;$^{(f)}$Institute of Applied Physics, Mohammed VI Polytechnic University, Ben Guerir; Morocco.\\
% ...
$^{45}$Physics Department, Southern Methodist University, Dallas TX; United States of America.\\
% ...
$^{49}$Fakult\"{a}t Physik , Technische Universit{\"a}t Dortmund, Dortmund; Germany.\\
% ...
$^{55}$II. Physikalisches Institut, Georg-August-Universit\"{a}t G\"ottingen, G\"ottingen; Germany.\\
% ...
$^{104}$CPPM, Aix-Marseille Universit\'e, CNRS/IN2P3, Marseille; France.\\
% ...
$^{119}$$^{(a)}$New York University Abu Dhabi, Abu Dhabi;$^{(b)}$United Arab Emirates University, Al Ain; United Arab Emirates.\\
% ...
$^{123}$Homer L. Dodge Department of Physics and Astronomy, University of Oklahoma, Norman OK; United States of America.\\
% ...
$^{157}$Raymond and Beverly Sackler School of Physics and Astronomy, Tel Aviv University, Tel Aviv; Israel.\\
% ...
$^{a}$ Also at Affiliated with an institute formerly covered by a cooperation agreement with CERN.\\
$^{b}$ Also at An-Najah National University, Nablus; Palestine.\\
$^{c}$ Also at Borough of Manhattan Community College, City University of New York, New York NY; United States of America.\\
$^{d}$ Also at Center for Interdisciplinary Research and Innovation (CIRI-AUTH), Thessaloniki; Greece.\\
$^{e}$ Also at Centre of Physics of the Universities of Minho and Porto (CF-UM-UP); Portugal.\\
$^{f}$ Also at CERN, Geneva; Switzerland.\\
% ...

\end{flushleft}


%

%
%

%

%
%
%
%

\end{document}